\documentclass[a4paper, onecolumn]{article}

\usepackage{fullpage}
\usepackage{graphicx}
\usepackage{amsmath}
\usepackage{amssymb}
\usepackage[margin=2cm]{caption}
\usepackage{mathtools}
\usepackage{subcaption}
\usepackage{psfrag}
\usepackage{tikz}
\usepackage{epstopdf}
\usepackage{glossaries}

\newcommand{\re}[1]{\mathrm{Re} \left\{#1\right\}}
\newcommand{\im}[1]{\mathrm{Im} \left\{#1\right\}}
\newcommand{\saeed}[1]{}
\newcommand{\saeedlong}[1]{}
\newcommand{\myvector}[1]{\textbf{#1}}
\newcommand{\eqn}[1]{(\ref{#1})}

\newcommand{\fig}[1]{Figure~\ref{#1}}
\newcommand{\doc}[1]{}
\newcommand{\tikzfolder}{}

\usetikzlibrary{positioning}
\usetikzlibrary{fit}

\tikzstyle{axis-line}=[thick, ->, >=stealth]
\tikzstyle{axis-tick}=[thin]
\tikzstyle{arrow} = [thick,->]

\tikzstyle{block}=[rectangle,black,draw, text centered, minimum height=0.7cm]
\tikzstyle{summul} = [draw, shape=circle, text centered,inner sep=0]

\newacronym{acOFDM}{OFDM}{Orthogonal Frequency Division Multiplexing}
\newacronym{acCMT}{CMT}{Cosine Modulated Multitone}
\newacronym{acSMT}{SMT}{Staggered Modulated Multitone}
\newacronym{acISI}{ISI}{Inter-Symbol Interference}
\newacronym{acICI}{ICI}{Inter-Channel Interference}

\title{An Introduction to OFDM-OQAM \\$2^\mathrm{nd}$ Edition}
\date{ July 27, 2017}
\author{Saeed Afrasiabi-Gorgani\\Ph.D. candidate \\ Technische Universit{\"a}t Berlin}

\begin{document}

\maketitle

The Filter Bank MultiCarrier (FBMC) modulation with Offset QAM (OQAM), also know as OFDM-OQAM, has been very well studied in, roughly speaking, the last 15 years. A rich literature exists on the principle itself, discrete-time version and implementation, equalization, extension to MIMO, etc. This text, written in a tutorial style, explains the idea of Offset QAM and why it works from a Communications Engineer point of view. The language is very simple and the math elementary. Students and researchers in universities or practitioners who are interested in gaining an intuition into the logic behind OFDM-OQAM will hopefully enjoy reading this text. On the other hand, those interested in a quick understanding of the idea, just enough to start using the waveform, can find far better material. 

In the first section, the general model for a multipulse modulation and its requirements are discussed. A multipulse modulation is a term referring to the general case where more than one pulse shape is used, of which the multicarrier modulation is a special case.

In the second section, the popular \gls{acOFDM} is explained. It is well-known that the \gls{acOFDM} lacks the features required for the emerging applications. This is mainly due to the rectangular pulse shape which slowly decays in the frequency domain. In the third section, it is shown that using other pulse shapes is not straightforward. In the fourth section, an old scheme from the 60s referred to as \gls{acCMT} is developed. Based on that, the fourth section explains the \gls{acSMT} scheme, which is the main objective of this text. OFDM-OQAM is only a small step away from SMT.


\subsection*{Acknowledgment}
This report was mostly written as a part of my Master's thesis, submitted in May 2014, under supervision of Professor Markku Renfors, in Tampere University of Technology, Finland. I have always been and will be grateful for having the experience of working with him.

There has been additions and modifications which have not been peer-reviewed.
\section{Multipulse modulation}

\begin{figure}[t]
\centering
%
%
%
\begin{tikzpicture}


\node[block,minimum width=2cm] (filter1) {$h_0(t)$};
\node[block,minimum width=2cm] (filter2) [below=of filter1] {$h_1(t)$};
\node[block,minimum width=2cm] (filter3) [below=of filter2, yshift=-2cm] {$h_{N-1}(t)$};

\node[circle, fill=black, below=of filter2, yshift=-0.5cm, inner sep=0pt, minimum width=0.1cm] {};
\node[circle, fill=black, below=of filter2, yshift=-0.7cm, inner sep=0pt, minimum width=0.1cm] {};
\node[circle, fill=black, below=of filter2, yshift=-0.9cm, inner sep=0pt, minimum width=0.1cm] {};

\foreach \x/\n in {1/0,2/1,3/N-1}
	\draw [<-] (filter\x.west) -- +(-0.8cm,0) node [anchor=south] {$A_{\n}$};

\node (adder) [summul,right of=filter2, xshift=2cm, yshift=-1cm] {\Large +};

\foreach \x in {1,2,3}
	\draw [->] (filter\x.east) -- +(1cm,0) -- (adder);

\draw [->] (adder.east) -- +(0.8cm,0) node [anchor=south] {$s(t)$};


\node[block,minimum width=2cm, right=of filter1, xshift=5.5cm] (rfilter1) {$h_0^\ast(-t)$};
\node[block,minimum width=2cm] (rfilter2) [below=of rfilter1] {$h_1^\ast(-t)$};
\node[block,minimum width=2cm] (rfilter3) [below=of rfilter2, yshift=-2cm] {$h_{N-1}^\ast(-t)$};

\node[circle, fill=black, below=of rfilter2, yshift=-0.5cm, inner sep=0pt, minimum width=0.1cm] {};
\node[circle, fill=black, below=of rfilter2, yshift=-0.7cm, inner sep=0pt, minimum width=0.1cm] {};
\node[circle, fill=black, below=of rfilter2, yshift=-0.9cm, inner sep=0pt, minimum width=0.1cm] {};

\foreach \x/\n in {1/0,2/1,3/N-1}
{
	\draw [->] (rfilter\x.east) -- +(0.5cm,0) node [anchor=north west] {$T$};
	\draw (rfilter\x.east) +(0.5cm,0) -- +(0.8cm,0.4cm);
	\draw [->] (rfilter\x.east) +(0.5cm,0.4) to [bend left=30] +(0.8cm,0cm);
	\draw (rfilter\x.east) +(0.5cm,0) +(0.8cm,0cm) -- +(1.4cm,0cm) [->] node [anchor=south] {$\hat{A}_{\n}$};
}

\node (input) [left of=rfilter2, xshift=-2cm,yshift=-1cm ] {} ;
\draw (input) -- +(1cm,0);
\foreach \x in {1,2,3}
	\draw (input) +(1cm,0) |- (rfilter\x.west) [->];

\foreach \x in {0.7,0.9,1.1}
	\node[circle, fill=black, left=of input, xshift=+\x cm, inner sep=0pt, minimum width=0.1cm] {};
	
\end{tikzpicture}
\caption{The general structure of the multipulse modulation.}
\label{fig:multipulse}
\end{figure}
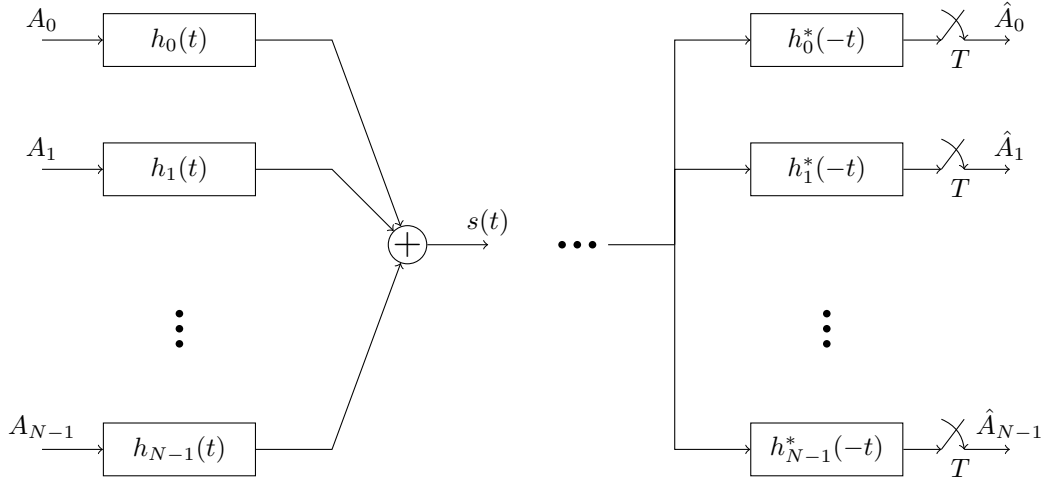

Consider a multipulse modulation of the form 
\begin{equation}
s(t)=\sum_{k=-\infty}^{+\infty} \sum_{n=0}^{N-1} A_{k,n} h_n(t-kT),
\label{eq:multipulse}
\end{equation}
where $A_{k,n}$ is the data symbol that modulates the $n^\mathrm{th}$ subcarrier in the $k^\mathrm{th}$ symbol interval. The modulation structure is illustrated in \fig{fig:multipulse}. In order to avoid \gls{acISI}, the set of pulses $h_n(t)$ for $n=0,\ldots,N-1$ must satisfy the ordinary Nyquist criterion
\begin{equation}
\left. h_n(t) \ast h_n^\ast(-t) \right|_{t=kT} = \delta_k, \quad n=0,\ldots,N-1,
\label{eq:ordinary-Nyquist-time}
\end{equation}
where $\delta_k$ is the Kronecker delta function and is defined as
\begin{equation}
\delta_k =
\begin{dcases}
1, \quad k=1\\
0,\quad k\neq1.
\end{dcases}
\end{equation}
Equation \eqn{eq:ordinary-Nyquist-time} is, in the frequency domain, equivalent to
\begin{equation}
\frac{1}{T} \sum_{m=-\infty}^{+\infty} |H_n(j(\omega+m\frac{2\pi}{T}))|^2 =1, \quad n=0,\ldots,N-1.
\label{eq:ordinary-Nyquist-freq}
\end{equation}
where $H_n(j\omega)$ is the Fourier transform of $h_n(t)$.
In order to avoid \gls{acICI}, the chosen pulses must satisfy the generalized Nyquist criterion \cite[p.~235]{lee}
\begin{equation}
\left. h_n(t) \ast h_l^\ast(-t) \right|_{t=kT} = \delta_k \delta_{l-n}, \quad n,l=0,\ldots,N-1,
\label{eq:generalized-Nyquist-time}
\end{equation}
or equivalently in the frequency domain
\begin{equation}
\frac{1}{T} \sum_{m=-\infty}^{+\infty} H_n(j(\omega+m\frac{2\pi}{T})) H_l^\ast(j(\omega+m\frac{2\pi}{T})) = \delta_{l-n},\quad n,l=0,\ldots,N-1.
\label{eq:generalized-Nyquist-freq}
\end{equation}
It is clear that the ordinary Nyquist criterion is the special case of its generalized version for $n=l$. 

A simple and straightforward way to design the pulses in \eqn{eq:multipulse} which satisfy the generalized Nyquist criterion is to isolate them in either the time or the frequency domain. The widely used OFDM scheme is based on isolation in the time domain, as will be explained in the next.

\saeedlong{introduce spectral efficiency, the default expectation for QAM, ...}

\saeedlong{With isolated pulses in the frequency domain, the minimum bandwidth is utilized only if the magnitude of the Fourier transform of the pulses is in a rectangular shape. This leads to pulses that are sinc functions in the time domain, which fade away too slowly. Hence, it is not possible to use a truncated version with a reasonable length. Therefore, isolation in the frequency domain is only possible with pulses taking more than the minimum bandwidth. Such schemes are referred to as filtered multitone (FMT).}

\section{Isolation in time, OFDM}
\label{sec:OFDM}
\saeedlong{rewrite this section. not sure about how much it must be elaborated, as it could distract the reader from the main path}

Time domain isolation is probably a natural choice. The rectangular pulse shape, and hence a sinc shape in the frequency domain, is a popular one. Such a pulse would then need infinite bandwidth as a sinc function is theoretically nonzero over $\mathbb{R}$. Therefore, the practical case is an approximation of such rectangular pulses. This scheme is referred to as the \acrfull{acOFDM}. In the sequel, the \gls{acOFDM} signal model is described based on the notation commonly used in the literature.

The continuous-time baseband signal model for a single symbol of the \gls{acOFDM} is
\begin{equation}
s(t)=\frac{1}{\sqrt{N}} \sum_{k=0}^{N-1} X_k e^{j 2\pi k f_\Delta t} \quad t\in[0,T],
\label{eq:OFDM-continuous}
\end{equation}
where $N$ is the number of subcarriers, the vector
\begin{equation}
\myvector{X}=[X_0,X_1,\ldots,X_{N-1}]^T
\end{equation}
denotes the data symbols, $T$ is the \gls{acOFDM} symbol duration and
\begin{equation}
f_\Delta=\frac{1}{T}
\end{equation}
is the frequency spacing of subcarriers. In this text, a data symbol is distinguished from an \gls{acOFDM} symbol explicitly unless it is clear from the context. Note that the pulses used in OFDM, according to \eqn{eq:multipulse}, are
\begin{equation}
h_n(t)= \mathrm{rect}(\frac{t-T/2}{T}) e^{j 2\pi \frac{n}{T} t}, \quad n=0,\ldots,N-1,
\label{eq:ofdm-pulse}
\end{equation}
where
\begin{equation}
\mathrm{rect}(t)=
\begin{cases}
1 & |t|\leq \frac{1}{2} \\
0 & \mathrm{elsewhere.}
\end{cases}
\end{equation}
The critically-sampled discrete-time signal model is accordingly obtained as
\begin{equation}
s[k]=s(k T_s)=\frac{1}{\sqrt{N}} \sum_{n=0}^{N-1} X_n e^{j 2\pi \frac{n}{T} kT_s},\quad k=0,1,\ldots,N-1,
\end{equation}
where the sampling frequency is $\frac{1}{T_s}=Nf_\Delta$. Therefore, $s[k]$ can be written as
\begin{equation}
s[k]=\frac{1}{\sqrt{N}} \sum_{n=0}^{N-1} X_n e^{j \frac{2\pi}{N} n k}, \quad k=0,1,\ldots,N-1.
\label{eq:OFDM-discrete}
\end{equation}
The scalar factor $\frac{1}{\sqrt{N}}$ is necessary to derive the statistical properties of the OFDM symbols.

It is interesting to observe that \eqn{eq:OFDM-discrete} is the expression for the Inverse Discrete Fourier Transform (IDFT). This makes it possible to use the very popular implementation of IDFT known as Fast Fourier Transform (FFT) to do the OFDM modulation. The scalar factor may vary among the different FFT implementations. 

\subsubsection*{Channel equalization}

The effect of a frequency-selective channel, caused by multipath propagation, on a signal can be modeled and illustrated by the effect of an FIR filter. The received signal is the sum of multiple replicas of the signal with different delays and attenuations. In other words, the symbols are mixed and the so-called \acrfull{acISI} happens. Thus the channel can be viewed as a filter in the transmitter-receiver chain, which is not taken into account when the pulse shape is designed. As a matter of fact, it is not possible to predict and account for the channel in advance. Therefore, the channel must be estimated and equalized. The problem is that equalizing a severely frequency-selective channel, which is the case in the modern applications, requires a high-order FIR filter. In addition, the equalization must be adaptive as the channel changes in time, which becomes more complex for such a filter. 

The main motivation for using the multicarrier modulation, particularly the \gls{acOFDM} as a widely used technique, is to avoid the complex channel equalization required for the single-carrier schemes. This facility requires that the channel functions as a subcarrier-wise multiplication of the data symbols with complex gains \cite{Pun:2007}. In other words, it is necessary that the channel response is flat in each subcarrier bandwidth.

In the OFDM, each subcarrier is spread over the entire available bandwidth. It takes a distance of several subcarriers until the so-called sidelobes drop to a negligible level. However, as the OFDM symbols are isolated in the time-domain, it is possible to change the linear convolution of the channel to circular convolution and establish the subcarrier-wise multiplication of the frequency response of the channel with the data symbols. This is done by inserting a Cyclic Prefix (CP) to the beginning of the OFDM symbol \cite[Chapter~8]{DSP-oppenheim}. In the receiver side, the CP is discarded before the demodulation.

In the schemes developed in the following sections, the pulse shapes are well-localized in the frequency domain. If the bandwidth of the subchannels is relatively small and the number of them is high, then each of them experiences an approximately flat channel response \cite{5753092}.

\section{Overlapping in time and frequency}
\label{sec:overlapping-failure}
A major drawback of the OFDM scheme is the heavy overlapping of the neighboring subchannels in the frequency domain. This drawback stands out most when it is desired to have the multicarrier system work in coexistence with other systems \cite{6692670}. It is often necessary to have a spectrally well-contained waveform, a property that the OFDM waveform mainly lacks.

Here we investigate the problem from the viewpoint of the uniform framework explained before. It is clear that pulses with mild behavior in the time domain have better spectral containment. For simplicity, and to keep the discussion close to the well-known schemes, pulses are allowed to overlap over only the adjacent subchannels. This corresponds to a 100\% or lower roll-off.

A first thought would suggest to modify the OFDM modulation only by changing the pulse shape from a $\mathrm{rect}(t/T)$ to a more well-behaved waveform $q(t)$, i.e.,
\begin{equation}
h_n(t)=q(t) e^{j \frac{2\pi}{T} n t},
\label{eq:modified-ofdm}
\end{equation}
where $q(t)$ is the basic pulse shape designed for a symbol duration of $T$. The frequency transform of each pulse is 
\begin{equation}
H_n(j\omega)=Q(j(\omega-n\frac{2\pi}{T})),
\end{equation}
where $Q(j\omega)$ is the Fourier transform of $q(t)$.
Applying the ordinary Nyquist criterion as in \eqn{eq:ordinary-Nyquist-freq}, $Q(j\omega)$ must satisfy
\begin{equation}
\frac{1}{T} \sum_{m=-\infty}^{+\infty} |Q(j(\omega-n\frac{2\pi}{T}+m\frac{2\pi}{T}))|^2 = 1,
\end{equation}
which can be rewritten via a change of variable as
\begin{equation}
\frac{1}{T} \sum_{m=-\infty}^{+\infty} |Q(j(\omega-m\frac{2\pi}{T}))|^2 = 1.
\end{equation}
This shows that $Q(j\omega)$ must be a Nyquist pulse, e.g., as the popular root-raised cosine pulse. Note that the ordinary Nyquist criterion restricts only the magnitude of~$Q(j\omega)$.

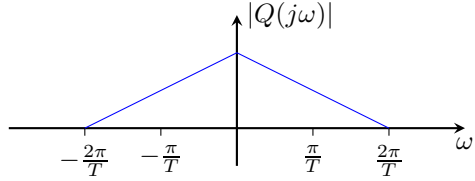
\begin{figure}[!t]%
\centering


\begin{tikzpicture}
\draw [axis-line] (-3,0) -- (3,0) node[below] {$\omega$};
\draw [axis-line] (0,-0.5) -- (0,1.5) node[right, black] {$|Q(j\omega)|$};
\draw [axis-tick] (-2,0) -- +(0,-0.1)node[below] {$-\frac{2\pi}{T}$};
\draw [axis-tick] (2,0) -- +(0,-0.1)  node[below] {$\frac{2\pi}{T}$};
\draw [axis-tick] (1,0) -- +(0,-0.1)  node[below] {$\frac{\pi}{T}$};
\draw [axis-tick] (-1,0) -- +(0,-0.1)  node[below] {$-\frac{\pi}{T}$};

\draw [blue] (-2,0) -- (0,1) -- (2,0);

\end{tikzpicture}

\caption{A sketch of the amplitude of the basic pulse shape in frequency domain, $Q(j\omega)$}%
\label{fig:pulse-sketch}%
\end{figure}

Following the assumption that $Q(j\omega)$ overlaps only over the adjacent subchannels, as sketched in \fig{fig:pulse-sketch}, the generalized Nyquist criterion as in \eqn{eq:generalized-Nyquist-freq} can be simplified. It is clear that the criterion is satisfied for $l-n>1$, as the more distant pulses never overlap theoretically. Hence,
\begin{equation}
\frac{1}{T} \sum_{m=-\infty}^{+\infty} H_n(j(\omega+m\frac{2\pi}{T})) H_{n+1}^\ast(j(\omega+m\frac{2\pi}{T})) = 0
\end{equation}
must be met. In terms of $Q(j\omega)$, and after a change of variable, we have
\begin{equation}
\frac{1}{T} \sum_{m=-\infty}^{+\infty} Q(j(\omega+m\frac{2\pi}{T})) Q^\ast(j(\omega+(m-1)\frac{2\pi}{T})) = 0.
\label{eq:generalized-nyquist-failure}
\end{equation}
Note that the choices of $l=n+1$ and $l=n-1$ are equivalent.

\begin{figure}[!t]%
\centering
%
%
\begin{tikzpicture}
\draw [axis-line] (-7,0) -- (7,0) node[below] {$\omega$};
\draw [axis-line] (0,-0.5) -- (0,3) node[right, black] {$|H_n(j\omega)|$};
\foreach \x/\n in {1/\!,2/2,3/3,4/4,5/5,6/6}
{
	\draw [axis-tick] (\x,0) -- +(0,-0.1) node[below] {\small $\frac{\n\pi}{T}$};
	\draw [axis-tick] (-\x,0) -- +(0,-0.1) node[below] {\small $-\frac{\n\pi}{T}$};
}

\draw [black, thick] (-2,0) -- +(2,1) node [below, yshift=-0.2cm] {\tiny $n\!=\!0$}-- +(4,0);
\draw [black, thick] (0,0) -- +(2,1) node [below, yshift=-0.2cm] {\tiny $n\!=\!1$} -- +(4,0);

\draw [blue] (-4,0) -- +(2,1) -- +(4,0);
\draw [blue] (-1.9,0) -- +(2,1) -- +(4,0);

\draw [red] (0.1,0) -- +(2,1) -- +(4,0);
\draw [red] (2,0) -- +(2,1) -- +(4,0);

\fill [draw=black, fill=black!30!white, rounded corners=0.03cm] (0,1.2) rectangle +(2,-0.15);

\fill [draw=blue, fill=blue!30!white, rounded corners=0.03cm] (-2,1.2) rectangle +(1.95,-0.15);

\fill [draw=red, fill=red!30!white, rounded corners=0.03cm] (2.05,1.2) rectangle +(2,-0.15);

\draw [red,->] (1,1.5) .. controls (1.5,2) and (2.5,2) .. (3,1.5) node[above, midway] {$m\!=\!-1$};
\draw [blue,->] (0.9,1.5) .. controls (0.5,2) and (-0.5,2) .. (-1,1.5) node[above, midway] {$m\!=\!+1$};

\end{tikzpicture}
%
%
%
\caption{The construction of the expression in \eqn{eq:generalized-nyquist-failure}. A hashed area indicates the interval where a product is nonzero.}%
\label{fig:nyquist-failure-case}%
\end{figure}
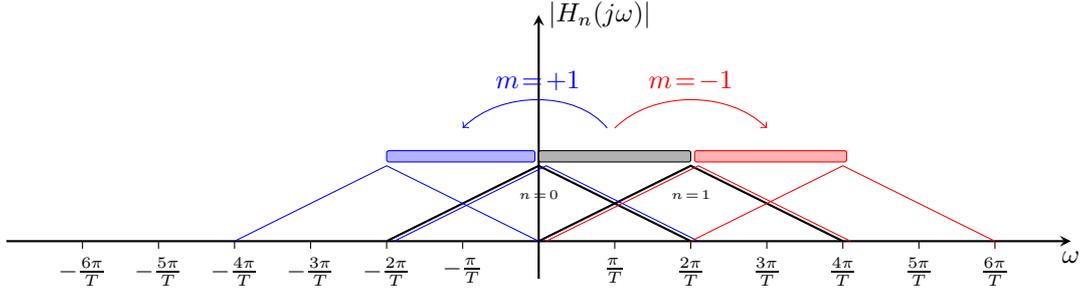

The expression in \eqn{eq:generalized-nyquist-failure} is clearly periodic over intervals of $\frac{2\pi}{T}$. Therefore, we can investigate the interval of $\omega\in[-\frac{\pi}{T},\frac{\pi}{T}]$. \fig{fig:nyquist-failure-case} reveals the contribution of each term to the summation in \eqn{eq:generalized-nyquist-failure}. The black curves refer to the adjacent pulses for $n=0$ and $n=1$. The shifted copies of the product of these pulses build the summation in \eqn{eq:generalized-nyquist-failure}. For simplicity, only the nonzero interval of the product is indicated by a gray hashed box. Focusing on $\omega\in[-\frac{\pi}{T},\frac{\pi}{T}]$, the figure shows all the terms which contribute and their nonzero intervals.

Therefore, it is clear from \fig{fig:nyquist-failure-case} that the summation in the desired interval can be simplified to 
\begin{equation}
Q(j\omega) Q^\ast(j(\omega-\frac{2\pi}{T})) + Q(j(\omega+\frac{2\pi}{T})) Q^\ast(j\omega) =0, \quad \omega\in[-\frac{\pi}{T},\frac{\pi}{T}],
\label{eq:generalized-nyquist-failure2}
\end{equation}
which includes the terms corresponding to $m=0$ and $m=1$. Furthermore, \fig{fig:nyquist-failure-case} shows that the nonzero intervals of the products in \eqn{eq:generalized-nyquist-failure} do not overlap. In other words, both of the terms in \eqn{eq:generalized-nyquist-failure2} must be zero. Furthermore, since they differ only by a frequency shift, the generalized Nyquist criterion finally reduces to
\begin{equation}
Q(j\omega) Q^\ast(j(\omega-\frac{2\pi}{T}))=0, \quad \omega\in[-\frac{\pi}{T},\frac{\pi}{T}].
\label{eq:generalized-nyquist-failure-final}
\end{equation}

It is clear that generally for a complex number $a$, $|a| e^{j\theta_a}=0$ if and only if $|a|=0$. Thus $|Q(j\omega)|$ cannot lead to \eqn{eq:generalized-nyquist-failure-final} as it would be absurd.
Therefore, it is not possible to satisfy the generalized Nyquist criterion by these pulses. In other words, following the general idea of the OFDM and changing the pulse shapes does not lead to a working scheme.

\section{Cosine Modulated Multitone (CMT)}

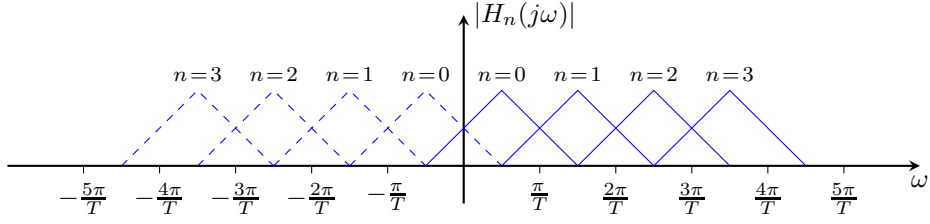
\begin{figure}[!t]%
\centering
%
%
\begin{tikzpicture}
\draw [axis-line] (-6,0) -- (6,0) node[below] {$\omega$};
\draw [axis-line] (0,-0.5) -- (0,2) node[right, black] {$|H_n(j\omega)|$};
\foreach \x/\n in {1/\!,2/2,3/3,4/4,5/5}
{
	\draw [axis-tick] (\x,0) -- +(0,-0.1) node[below] {$\frac{\n\pi}{T}$};
	\draw [axis-tick] (-\x,0) -- +(0,-0.1) node[below] {$-\frac{\n\pi}{T}$};
}

\foreach \x/\n in {-4.5/3,-3.5/2,-2.5/1,-1.5/0}
	\draw [dashed,blue] (\x,0) -- +(1,1) node[above,black] {\footnotesize $n\!=\!\n$} -- +(2,0);
\foreach \x/\n in {-0.5/0,0.5/1,1.5/2,2.5/3}
	\draw  [blue](\x,0) -- +(1,1) node[above,black] {\footnotesize $n\!=\!\n$} -- +(2,0);

\end{tikzpicture}
%
%
%
\caption{A frequency domain illustration of the pulses in the CMT scheme. The dashed curves indicate the negative frequency part of the double-sided pulses.}%
\label{fig:CMT-pulses}%
\end{figure}

There is a working modulation scheme first developed by Chang \cite{Chang}, which is the basis for the so-called \acrfull{acCMT} scheme. The pulses $h_n(t)$, as in \eqn{eq:multipulse}, are defined as
\begin{equation}
h_n(t)=q(t) \cos{((n+\frac{1}{2})\frac{\pi}{T} t)}, \quad n=0,1,\ldots,N-1,
\label{eq:CMT-pulses-time}
\end{equation}
where $q(t)$ is a Nyquist pulse for a symbol duration of $2T$. In the frequency domain, they are
\begin{equation}
H_n(j\omega)=\frac{1}{2}\left[Q(j(\omega+\frac{\pi}{2T}+n\frac{\pi}{T}))+Q(j(\omega-\frac{\pi}{2T}-n\frac{\pi}{T}))\right] , \quad n=0,1,\ldots,N-1.
\label{eq:CMT-pulses-freq}
\end{equation}
\fig{fig:CMT-pulses} illustrates the arrangement of the pulses in the frequency domain. The minimum bandwidth required for $q(t)$ is $\frac{\pi}{T}$. Repeating the assumption of 100\% roll-off, it needs twice the minimum bandwidth, $\frac{2\pi}{T}$. Thus the frequency separation is $\frac{\pi}{T}$. However, the key difference here is that the pulses are real and double-sided in frequency. If real data symbols are used, the symmetry of the pulses allows the application of the Vestigial Side Band (VSB) modulation. 

It is clear from \fig{fig:CMT-pulses} that $h_n(t), n=1, 2,\ldots, N$ satisfies the ordinary Nyquist criterion for the symbol duration $T$ provided that $q(t)$ does so for $2T$. \saeed{maybe not that clear} Here the role of the extra frequency shift by $\frac{\pi}{2T}$ and then the modulation to $-\frac{n\pi}{T}$ and $\frac{n\pi}{T}$ becomes clear. 

The ordinary Nyquist criterion sets constraints on $A(\omega)=|Q(j\omega)|$. Applying the generalized criterion will then reveal the requirement for the phase response $\mathrm{\theta}(\omega)=\mathrm{arg}\ Q(j\omega)$ as will be explained shortly. Since the pulses overlap only over the adjacent ones, then it is enough to consider the cases $l=n+1$ or $l=n-1$. It is clear from \eqn{eq:generalized-Nyquist-freq} that these two cases are equivalent. Rewriting \eqn{eq:generalized-Nyquist-freq} for $l=n+1$, we have
\begin{equation}
\frac{1}{T} \sum_{m=-\infty}^{+\infty} H_n(j(\omega+m\frac{2\pi}{T})) H_{n+1}^\ast(j(\omega+m\frac{2\pi}{T})) = 0.
\label{eq:CMT-generalized-first-step}
\end{equation}
Substituting \eqn{eq:CMT-pulses-freq} and after some manipulation, it follows as
\begin{align}
\frac{1}{T} \sum_{m=-\infty}^{+\infty} 
[ Q(j(\omega-(2n+1)\frac{\pi}{2T}+ m\frac{2\pi}{T})) &+  Q(j(\omega+(2n+1)\frac{\pi}{2T}+ m\frac{2\pi}{T})) ]\times \nonumber \\ 
[ Q^\ast(j(\omega-(2n+3)\frac{\pi}{2T}+ m\frac{2\pi}{T}))&+  Q^\ast(j(\omega+(2n+3)\frac{\pi}{2T}+ m\frac{2\pi}{T})) ]  = 0.
\end{align}
Two of the four product terms vanish as they do not overlap. Therefore, we have
\begin{align}
\frac{1}{T} \sum_{m=-\infty}^{+\infty} 
Q(j(\omega-(2n+1)\frac{\pi}{2T}+ m\frac{2\pi}{T})) & Q^\ast(j(\omega-(2n+3)\frac{\pi}{2T}+ m\frac{2\pi}{T}))+\nonumber \\
Q(j(\omega+(2n+1)\frac{\pi}{2T}+ m\frac{2\pi}{T})) & Q^\ast(j(\omega+(2n+3)\frac{\pi}{2T}+ m\frac{2\pi}{T})) = 0.
\label{eq:CMT-nyquist-general}
\end{align}
Since this expression is $2\pi$-periodic, its evaluation  can be made easier by inspecting it over only $\omega\in[-\frac{\pi}{T},\frac{\pi}{T}]$.

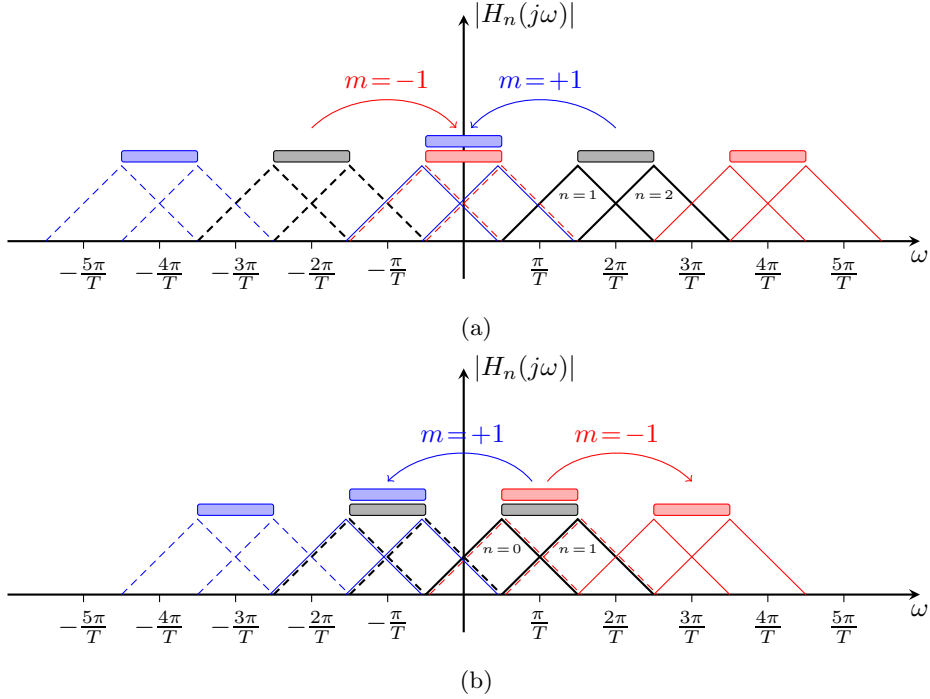
\begin{figure}[!t]%
\centering
\begin{subfigure}{\textwidth}
\centering
%
%
\begin{tikzpicture}
\draw [axis-line] (-6,0) -- (6,0) node[below] {$\omega$};
\draw [axis-line] (0,-0.5) -- (0,3) node[right, black] {$|H_n(j\omega)|$};
\foreach \x/\n in {1/\!,2/2,3/3,4/4,5/5}
{
	\draw [axis-tick] (\x,0) -- +(0,-0.1) node[below] {$\frac{\n\pi}{T}$};
	\draw [axis-tick] (-\x,0) -- +(0,-0.1) node[below] {$-\frac{\n\pi}{T}$};
}

\draw [black, thick] (0.5,0) -- +(1,1) node [below, yshift=-0.2cm] {\tiny $n\!=\!1$}-- +(2,0);
\draw [black, thick] (1.5,0) -- +(1,1) node [below, yshift=-0.2cm] {\tiny $n\!=\!2$} -- +(2,0);
\draw [black, thick, densely dashed] (-3.5,0) -- +(1,1) -- +(2,0);
\draw [black, thick, densely dashed] (-2.5,0) -- +(1,1) -- +(2,0);

\draw [blue] (-1.55,0) -- +(1,1) -- +(2,0);
\draw [blue] (-.55,0) -- +(1,1) -- +(2,0);
\draw [blue,densely dashed] (-4.5,0) -- +(1,1) -- +(2,0);
\draw [blue,densely dashed] (-5.5,0) -- +(1,1) -- +(2,0);

\draw [red] (2.5,0) -- +(1,1) -- +(2,0);
\draw [red] (3.5,0) -- +(1,1) -- +(2,0);
\draw [red,densely dashed] (-1.5,0) -- +(1,1) -- +(2,0);
\draw [red,densely dashed] (-0.5,0) -- +(1,1) -- +(2,0);

\filldraw [draw=black, fill=black!30!white, rounded corners=0.03cm] (1.5,1.2) rectangle +(1,-0.15);
\filldraw [draw=black, fill=black!30!white, rounded corners=0.03cm] (-2.5,1.2) rectangle +(1,-0.15);

\filldraw [draw=blue, fill=blue!30!white, rounded corners=0.03cm] (-0.5,1.4) rectangle +(1,-0.15);
\filldraw [draw=blue, fill=blue!30!white, rounded corners=0.03cm] (-4.5,1.2) rectangle +(1,-0.15);

\filldraw [draw=red, fill=red!30!white, rounded corners=0.03cm] (3.5,1.2) rectangle +(1,-0.15);
\filldraw [draw=red, fill=red!30!white, rounded corners=0.03cm] (-0.5,1.2) rectangle +(1,-0.15);

\draw [red,->] (-2,1.5) .. controls (-1.5,2) and (-0.5,2) .. (-0.1,1.5) node[above, midway] {$m\!=\!-1$};
\draw [blue,->] (2,1.5) .. controls (1.5,2) and (0.5,2) .. (0.1,1.5) node[above, midway] {$m\!=\!+1$};

\end{tikzpicture}
%
%
%
\caption{}
\label{fig:CMT-nyquist-h1}%
\end{subfigure}
\begin{subfigure}{\textwidth}
\centering
%
%
\begin{tikzpicture}
\draw [axis-line] (-6,0) -- (6,0) node[below] {$\omega$};
\draw [axis-line] (0,-0.5) -- (0,3) node[right, black] {$|H_n(j\omega)|$};
\foreach \x/\n in {1/\!,2/2,3/3,4/4,5/5}
{
	\draw [axis-tick] (\x,0) -- +(0,-0.1) node[below] {$\frac{\n\pi}{T}$};
	\draw [axis-tick] (-\x,0) -- +(0,-0.1) node[below] {$-\frac{\n\pi}{T}$};
}

\draw [black, thick] (-0.5,0) -- +(1,1) node [below, yshift=-0.2cm] {\tiny $n\!=\!0$}-- +(2,0);
\draw [black, thick] (0.5,0) -- +(1,1) node [below, yshift=-0.2cm] {\tiny $n\!=\!1$} -- +(2,0);
\draw [black, thick, densely dashed] (-2.5,0) -- +(1,1) -- +(2,0);
\draw [black, thick, densely dashed] (-1.5,0) -- +(1,1) -- +(2,0);

\draw [blue] (-2.55,0) -- +(1,1) -- +(2,0);
\draw [blue] (-1.55,0) -- +(1,1) -- +(2,0);
\draw [blue,densely dashed] (-3.5,0) -- +(1,1) -- +(2,0);
\draw [blue,densely dashed] (-4.5,0) -- +(1,1) -- +(2,0);

\draw [red] (1.5,0) -- +(1,1) -- +(2,0);
\draw [red] (2.5,0) -- +(1,1) -- +(2,0);
\draw [red,densely dashed] (-0.45,0) -- +(1,1) -- +(2,0);
\draw [red,densely dashed] (0.55,0) -- +(1,1) -- +(2,0);

\filldraw [draw=black, fill=black!30!white, rounded corners=0.03cm] (0.5,1.2) rectangle +(1,-0.15);
\filldraw [draw=black, fill=black!30!white, rounded corners=0.03cm] (-1.5,1.2) rectangle +(1,-0.15);

\filldraw [draw=blue, fill=blue!30!white, rounded corners=0.03cm] (-1.5,1.4) rectangle +(1,-0.15);
\filldraw [draw=blue, fill=blue!30!white, rounded corners=0.03cm] (-3.5,1.2) rectangle +(1,-0.15);

\filldraw [draw=red, fill=red!30!white, rounded corners=0.03cm] (2.5,1.2) rectangle +(1,-0.15);
\filldraw [draw=red, fill=red!30!white, rounded corners=0.03cm] (0.5,1.4) rectangle +(1,-0.15);

\draw [red,->] (1.1,1.5) .. controls (1.5,2) and (2.5,2) .. (3,1.5) node[above, midway] {$m\!=\!-1$};
\draw [blue,->] (0.9,1.5) .. controls (0.5,2) and (-0.5,2) .. (-1,1.5) node[above, midway] {$m\!=\!+1$};

\end{tikzpicture}
%
%
%
\caption{}
\label{fig:CMT-nyquist-h0}%
\end{subfigure}
\caption{Illustration of the procedure taken to reach at \eqn{eq:CMT-nyquist-general} for (a) $n=1$ and (b) $n=0$. The blue, red and black components indicate the $m=1$, $m=-1$ and $m=0$ terms, respectively. Dashed lines show the negative frequency part of the pulses. The hashed areas indicate nonzero intervals.}
\label{fig:CMT-nyquist}
\end{figure}

The procedure to derive \eqn{eq:CMT-nyquist-general} is visualized in \fig{fig:CMT-nyquist-h1} for $n=1$ and in \fig{fig:CMT-nyquist-h0} for $n=0$. \saeedlong{The figures show slightly different situation for $n=1$ and $n=0$. In fact, for all even $n$ we get something like that of $n=0$. And for all odd $n$, something similar to $n=1$. I just don't know if there's a nice interpretation behind this.} Here we proceed with the case of $n=1$, as an example. It is clear that in \eqn{eq:CMT-nyquist-general} the terms referring to $m=1$ and $m=-1$ partly fall into the interval of $\omega\in[-\frac{\pi}{T},\frac{\pi}{T}]$. Therefore, removing the remaining frequency shifts that fall outside the interval, we have
\begin{equation}
Q(j(\omega-\frac{\pi}{2T})) Q^\ast(j(\omega+\frac{\pi}{2T}))+Q(j(\omega+\frac{\pi}{2T}))  Q^\ast(j(\omega-\frac{\pi}{2T})) = 0,
\label{eq:CMT-nyquist-before-real}
\end{equation}
which is, since $\alpha+\alpha^\ast=2\re{\alpha}$, equivalent to
\begin{equation}
2\re{Q(j(\omega-\frac{\pi}{2T})) Q^\ast(j(\omega+\frac{\pi}{2T}))} = 0.
\label{eq:CMT-nyquist-real}
\end{equation}
Rewriting it in terms of the magnitude and phase of $Q(j\omega)$, we can find a constraint on the phase as
\begin{equation}
\cos(\mathrm{\theta}(\omega-\frac{\pi}{2T})+\mathrm{\theta}(-\omega-\frac{\pi}{2T}))=0, \quad \omega\in[-\frac{\pi}{T},\frac{\pi}{T}],
\end{equation}
or equivalently
\begin{equation}
\mathrm{\theta}(\omega-\frac{\pi}{2T})+\mathrm{\theta}(-\omega-\frac{\pi}{2T})=p\frac{\pi}{2},\quad p\in\mathbb{Z} \quad \omega\in[-\frac{\pi}{T},\frac{\pi}{T}].
\label{eq:CMT-phase-constraint}
\end{equation}
This implies a constraint on the phase of $q(t)$, which is easily achievable as will be explained shortly. Notice that this is exactly where the approach taken in section \ref{sec:overlapping-failure} failed. A comparison between  \fig{fig:nyquist-failure-case} and \fig{fig:CMT-nyquist} is illustrative in understanding the chain of events.

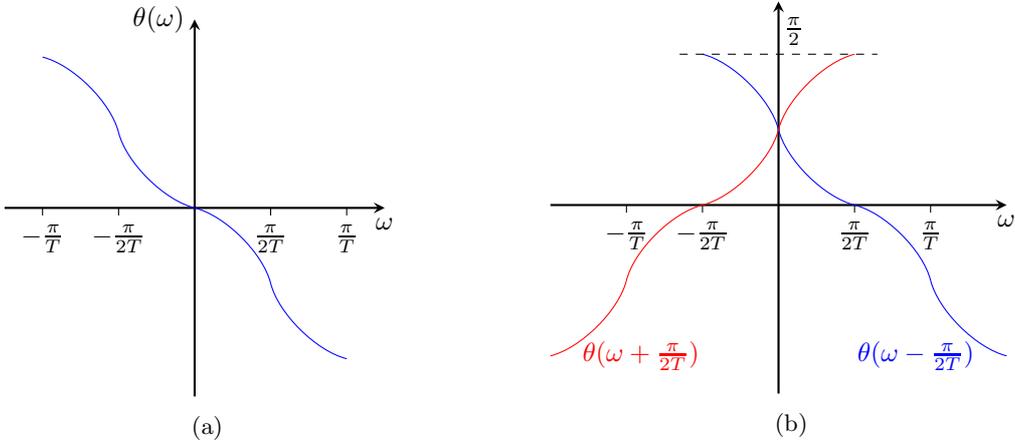
\begin{figure}[!t]%
\centering
\begin{subfigure}{0.45\textwidth}
\centering
%
%
\begin{tikzpicture}[scale=1]
\draw [axis-line] (-2.5,0) -- (2.5,0) node[below] {$\omega$};
\draw [axis-line] (0,-2.5) -- (0,2.5) node[anchor=east] {$\theta(\omega)$};

\draw [axis-tick] (1,0) -- +(0,-0.1) node[below] {$\frac{\pi}{2T}$};
\draw [axis-tick] (2,0) -- +(0,-0.1) node[below] {$\frac{\pi}{T}$};

\draw [axis-tick] (-1,0) -- +(0,-0.1) node[below] {$-\frac{\pi}{2T}$};
\draw [axis-tick] (-2,0) -- +(0,-0.1) node[below] {$-\frac{\pi}{T}$};

\draw [blue] (-2,2) .. controls +(0.4,-0.1) and +(-0.1,0.4) .. +(1,-1);
\draw [blue] (-1,1) .. controls +(0.1,-0.4) and +(-0.4,0.1) .. +(1,-1);	\draw [blue] (0,0) .. controls +(0.4,-0.1) and +(-0.1,0.4) .. +(1,-1);	\draw [blue] (1,-1) .. controls +(0.1,-0.4) and +(-0.4,0.1) .. +(1,-1);	

\end{tikzpicture}
%
%
%
\caption{}
\end{subfigure}
\begin{subfigure}{0.50\textwidth}
\centering
%
%
\begin{tikzpicture}[scale=1]
\draw [axis-line] (-3,0) -- (3,0) node[below] {$\omega$};
\draw [axis-line] (0,-2.5) -- (0,2.7) ;

\draw [axis-tick] (1,0) -- +(0,-0.1) node[below] {$\frac{\pi}{2T}$};
\draw [axis-tick] (2,0) -- +(0,-0.1) node[below] {$\frac{\pi}{T}$};

\draw [axis-tick] (-1,0) -- +(0,-0.1) node[below] {$-\frac{\pi}{2T}$};
\draw [axis-tick] (-2,0) -- +(0,-0.1) node[below] {$-\frac{\pi}{T}$};

\draw [blue] (-1,2) .. controls +(0.4,-0.1) and +(-0.1,0.4) .. +(1,-1);
\draw [blue] (0,1) .. controls +(0.1,-0.4) and +(-0.4,0.1) .. +(1,-1);	\draw [blue] (1,0) .. controls +(0.4,-0.1) and +(-0.1,0.4) .. +(1,-1);	\draw [blue] (2,-1) .. controls +(0.1,-0.4) and +(-0.4,0.1) .. +(1,-1) node [anchor= east, xshift=-0.3cm] {$\theta(\omega-\frac{\pi}{2T})$};	

\draw [red] (1,2) .. controls +(-0.4,-0.1) and +(0.1,0.4) .. +(-1,-1);
\draw [red] (0,1) .. controls +(-0.1,-0.4) and +(0.4,0.1) .. +(-1,-1);	
\draw [red] (-1,0) .. controls +(-0.4,-0.1) and +(0.1,0.4) .. +(-1,-1);
\draw [red] (-2,-1) .. controls +(-0.1,-0.4) and +(0.4,0.1) .. +(-1,-1) node [anchor= west, xshift=0.3cm] {$\theta(\omega+\frac{\pi}{2T})$};

\draw [black, dashed] (-1.3,2) -- (1.3,2) node [anchor=south, midway, xshift=0.2cm] {$\frac{\pi}{2}$};

\end{tikzpicture}
%
%
%
\caption{}
\end{subfigure}
\caption{An illustration of the phase constraint \eqn{eq:CMT-phase-constraint} imposed by the generalized Nyquist criterion. (a) A generic sketch of $\mathrm{\theta}(\omega)$, the phase response of the basic pulse $q(t)$. (b) The terms in \eqn{eq:CMT-phase-constraint}.}
\label{fig:CMT-phase-constraint}
\end{figure}

The situation with the phase requirement is depicted in \fig{fig:CMT-phase-constraint}. Knowing that $\mathrm{\theta}(\omega)$ is odd, it can be seen from the figure that a phase symmetry about $\frac{\pi}{2T}$ is required. A practical consideration is to use a zero-phase or symmetric $q(t)$, where $\mathrm{\theta}(\omega)=0$. In this special case, the phase requirement can be met by simply adding a phase shift of $\frac{\pi}{2}$ to every other pulse. This can be readily justified by considering a multiplicative term $e^{\pm j\pi/2}$ added to either $H_n(j\omega)$ or $H_{n+1}(j\omega)$ in \eqn{eq:CMT-generalized-first-step}. 

Since the developed baseband signal is real for real data symbols, hence symmetric in the frequency domain, only a single side-band is sufficient for the modulation. This way, it provides a symbol rate of $\frac{1}{T}$ for an aggregate bandwidth of about $\frac{N}{2T}$. Considering that a QAM symbol could be viewed as two real symbols, this is equal to the bandwidth efficiency of the OFDM scheme.

\saeedlong{We have a set of pulses that satisfy Nyquist criteria. So we must be able to use them for QAM symbols as well. If so, the baseband signal is no longer real and symmetric. However it takes the expected BW of $\frac{N}{T}$ for a symbol rate of $\frac{1}{T}$. Then DSB modulation can be used. The question is that why SMT or OQAM/OFDM are based on CMT? Or why is it necessary to go through all the staggering and make it complicated?}

\begin{figure}[!t]%
\centering
\begin{subfigure}{0.7\textwidth}
\centering
%
%
%
\begin{tikzpicture}

\node[block] (filter1) {$q(t) e^{j\frac{\pi}{2T}t}$};
\node[block] (filter2) [below=of filter1] {$q(t) e^{j\frac{\pi}{2T}t}$};
\node[block] (filter3) [below=of filter2, yshift=-1.2cm] {$q(t) e^{j\frac{\pi}{2T}t}$};

\node[circle, fill=black, below=of filter2, yshift=0cm, inner sep=0pt, minimum width=0.1cm] {};
\node[circle, fill=black, below=of filter2, yshift=-0.2cm, inner sep=0pt, minimum width=0.1cm] {};
\node[circle, fill=black, below=of filter2, yshift=-0.4cm, inner sep=0pt, minimum width=0.1cm] {};

\foreach \x in {2,3}
	\node (mul\x) [summul,right of=filter\x, xshift=1cm] {\large $\times$};
	
\foreach \x in {2,3}
	\draw [->] (filter\x) -- (mul\x);
	
\foreach \x/\n in {1/0,2/1,3/N-1}
	\draw [<-] (filter\x.west) -- +(-0.8cm,0) node [anchor=south] {$A_{\n}$};

\draw [<-] (mul2) -- +(0cm,0.8cm) node [anchor=south west] {$e^{j \frac{\pi}{T}t}$};
\draw [<-] (mul3) -- +(0cm,0.8cm) node [anchor=south west] {$e^{j (N-1)\frac{\pi}{T}t}$};

\node (adder) [summul,right of=mul2, xshift=1.5cm, yshift=-1.3cm] {\Large +};

\draw [->] (filter1) -- +(3cm,0) -- (adder);
\foreach \x in {2,3}
	\draw [->] (mul\x) -- +(1cm,0) -- (adder);

\draw [->] (adder.east) -- +(0.8cm,0) node [anchor=south] {$s(t)$};

\end{tikzpicture}
\caption{}
\label{fig:CMT-modulator}
\end{subfigure}
\begin{subfigure}{0.7\textwidth}
\centering
%
%
%
\begin{tikzpicture}

\node[block] (filter1) {$q(-t) e^{-j\frac{\pi}{2T}t}$};
\node[block] (filter2) [below=of filter1] {$q(-t) e^{-j\frac{\pi}{2T}t}$};
\node[block] (filter3) [below=of filter2, yshift=-1.2cm] {$q(-t) e^{-j\frac{\pi}{2T}t}$};

\node[circle, fill=black, below=of filter2, yshift=-0.0cm, inner sep=0pt, minimum width=0.1cm] {};
\node[circle, fill=black, below=of filter2, yshift=-0.2cm, inner sep=0pt, minimum width=0.1cm] {};
\node[circle, fill=black, below=of filter2, yshift=-0.4cm, inner sep=0pt, minimum width=0.1cm] {};

\foreach \x in {1,2,3}
	\node(Re\x) [block, right of=filter\x, xshift=1cm] {Re};
	
\foreach \x in {1,2,3}
	\draw [->] (filter\x) -- (Re\x);

\foreach \x in {2,3}
	\node (mul\x) [summul,left of=filter\x, xshift=-1.2cm] {\large $\times$};

\foreach \x in {2,3}
	\draw [<-] (filter\x) -- (mul\x);
	
\draw [<-] (mul2) -- +(0cm,0.8cm) node [anchor=south] {$e^{-j \frac{\pi}{T}t}$};
\draw [<-] (mul3) -- +(0cm,0.8cm) node [anchor=south] {$e^{-j (N-1)\frac{\pi}{T}t}$};

\foreach \x/\n in {1/0,2/1,3/N-1}
{
	\draw [->] (Re\x.east) -- +(0.5cm,0) node [anchor=north west] {$T$};
	\draw (Re\x.east) +(0.5cm,0) -- +(0.8cm,0.4cm);
	\draw [->] (Re\x.east) +(0.5cm,0.4) to [bend left=30] +(0.8cm,0cm);
	\draw (Re\x.east) +(0.5cm,0) +(0.8cm,0cm) -- +(1.4cm,0cm) [->] node [anchor=south] {$\hat{A}_{\n}$};
}

\node (input) [left of=filter2, xshift=-4cm,yshift=-1cm ,label=above:$s(t)$] {} ;
\draw (input) -- +(1cm,0);
\draw (input) +(1cm,0) |- (filter1.west) [->];
\draw (input) +(1cm,0) |- (mul2.west) [->];
\draw (input) +(1cm,0) |- (mul3.west) [->];

\end{tikzpicture}
\caption{}
\label{fig:CMT-demodulator}
\end{subfigure}
\caption{The CMT (a) modulator and (b) demodulator. Note the real operation after each filter.}%
\label{fig:CMT-mod-demod}%
\end{figure}
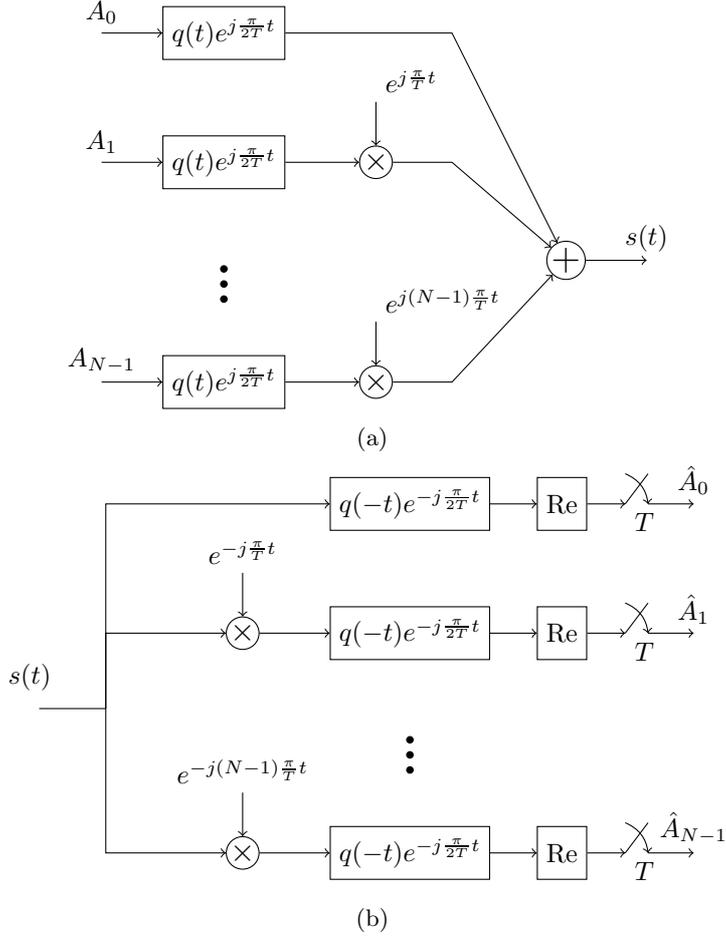

In order to use a single side-band, we can construct it directly without actually doing the required filtering in modulations such as VSB or SSB. This would modify the pulses to
\begin{equation}
h_n(t)=(q(t) e^{j \frac{\pi}{2T} t}) e^{j n \frac{\pi}{T} t}, \quad n=0,1,\ldots,N-1.
\label{eq:CMT-pulses-time-single-side}
\end{equation}
The rest of the operations in such a modulator are as usual, as depicted in \fig{fig:CMT-modulator}. The receiver, however, is slightly different. After demodulation to the baseband, the real part of the signal is obtained to form a double side-band signal. The block diagram of the receiver is depicted in \fig{fig:CMT-demodulator}. \saeed{elaborate?}

It is now straightforward to use the CMT scheme for the QAM symbols, by splitting each into two consecutive real symbols. These real symbols are defined as
\begin{align}
A'_{2k,n}&=A_{k,n}^R=\re{A_{k,n}}, \nonumber \\
A'_{2k+1,n}&=A_{k,n}^I=\im{A_{k,n}},
\end{align} 
where $A_{k,n}$ is the QAM symbol of time index $k$ and subcarrier $n$. With the notation developed so far, such a scheme would need $2T$ for each QAM symbol. In order to adjust the parameters to the typical ones of a QAM modulation, such that the QAM symbol interval is $T$, consider a basic pulse $q'(t)$ which satisfies ordinary Nyquist criterion for the symbol duration of $\frac{T}{2}$. Then the minimum bandwidth of the pulse, hence the required subcarrier spacing, is $\frac{2\pi}{T}$. Therefore, we can rewrite \eqn{eq:CMT-pulses-time} as
\begin{equation}
h'_n(t)=q'(t) \cos{((n+\frac{1}{2})\frac{2\pi}{T} t)}, \quad n=0,1,\ldots,N-1,
\end{equation}
or for the construction of the single side-band signal in the baseband,
\begin{equation}
h'_n(t)=q'(t) e^{j\frac{\pi}{T}t} e^{j n\frac{2\pi}{T}t}, \quad n=0,1,\ldots,N-1.
\end{equation}
\fig{fig:CMT-QAM-mod-demod} shows the block diagram representation of this scheme. In the diagram, a zero-phase pulse waveform is used and the generalized Nyquist criterion is satisfied by $\frac{\pi}{2}$ phase difference between the adjacent subchannels.


\section{Staggered Modulated Multitone (SMT)}
\label{sec:smt}
\fig{fig:CMT-phase} shows an interpretation of the phase rotation $e^{j\frac{\pi}{2T}t}$ applied to the basic pulse $q(t)$ as in \eqn{eq:CMT-pulses-time-single-side}. It it geometrically clear that this phase rotation matches the pulse duration of $T$ to create a phase difference of $\frac{\pi}{2}$ between the adjacent pulses. This can also be obtained mathematically by substituting \eqn{eq:CMT-pulses-time-single-side} in \eqn{eq:multipulse}
\begin{align}
s(t)&=\sum_{k=-\infty}^{+\infty} \sum_{n=0}^{N-1} 
A'_{k,n} q(t-kT) e^{j \frac{\pi}{2T}(t-kT)} e^{j n\frac{\pi}{T} (t-kT)} \nonumber \\
&=  e^{j \frac{\pi}{2T}t} \sum_{k=-\infty}^{+\infty} \sum_{n=0}^{N-1} A'_{k,n} q(t-kT) (-j)^k e^{j n\frac{\pi}{T} (t-kT)}.
\label{eq:CMT-phase-interp}
\end{align}
The term $(-j)^k$, which depends only on the time index $k$, implies the mentioned phase shift.

\begin{figure}[!th]%
\centering
\begin{subfigure}{\textwidth}
\centering
%
%
%
\begin{tikzpicture}

\node[block] (filter1) {$q'(t) e^{j\frac{\pi}{T}t}$};
\node[block] (filter2) [below=of filter1] {$q'(t) e^{j\frac{\pi}{T}t}$};
\node[block] (filter3) [below=of filter2] {$q'(t) e^{j\frac{\pi}{T}t}$};
\node[block] (filter4) [below=of filter3, yshift=-1.2cm] {$q'(t) e^{j\frac{\pi}{T}t}$};

\node[circle, fill=black, below=of filter3, yshift=-0.0cm, inner sep=0pt, minimum width=0.1cm] {};
\node[circle, fill=black, below=of filter3, yshift=-0.2cm, inner sep=0pt, minimum width=0.1cm] {};
\node[circle, fill=black, below=of filter3, yshift=-0.4cm, inner sep=0pt, minimum width=0.1cm] {};

\node (phasemul2) [summul,right of=filter2, xshift=1cm] {\large $\times$};
\node (phasemul4) [summul,right of=filter4, xshift=1cm] {\large $\times$};

\foreach \x in {2,3,4}
	\node (mul\x) [summul,right of=filter\x, xshift=2.3cm] {\large $\times$};
	
\draw [->] (filter2) -- (phasemul2);
\draw [->] (filter3) -- (mul3);
\draw [->] (filter4) -- (phasemul4);
\draw [->] (phasemul2) -- (mul2);
\draw [->] (phasemul4) -- (mul4);
	
\foreach \x/\n in {1/0,2/1,3/2, 4/(N-1)}
{
	\draw [<-] (filter\x.west) -- +(-0.8cm,0) node [anchor=south] {$A^I_{k,\n}$}
	-- +(-0.8cm,0.1cm);
	\draw (filter\x.west) +(-0.8cm,0) -- +(-2.2cm,0) node [anchor=south] {$A^R_{k,\n}$}
	-- +(-2.2cm,0.1cm);
}

\draw [<-] (mul2) -- +(0cm,0.8cm) node [anchor=south west] {$e^{j \frac{2\pi}{T}t}$};
\draw [<-] (mul3) -- +(0cm,0.8cm) node [anchor=south west] {$e^{j 2\frac{2\pi}{T}t}$};
\draw [<-] (mul4) -- +(0cm,0.8cm) node [anchor=south] {$e^{j (N-1)\frac{2\pi}{T}t}$};

\draw [<-] (phasemul2) -- +(0cm,0.8cm) node [anchor=south] {$e^{j \frac{\pi}{2}}$};
\draw [<-] (phasemul4) -- +(0cm,0.8cm) node [anchor=south] {$e^{j \frac{\pi}{2}}$};

\node (adder) [summul,right of=mul2, xshift=2cm, yshift=-1.3cm] {\Large +};

\draw [->] (filter1) -- +(4cm,0) -- (adder);
\foreach \x in {2,3,4}
	\draw [->] (mul\x) -- +(1cm,0) -- (adder);

\draw [->] (adder.east) -- +(0.8cm,0) node [anchor=south] {$s(t)$};

\draw [<-,thick,blue] (filter4.north west) +(0,0.7cm) node[anchor=north] {$t$} -- +(-2.7cm,0.7cm);
\draw [blue] (filter4.north west) +(-0.8cm,0.7cm) -- +(-0.8cm,0.8cm);
\draw [blue] (filter4.north west) +(-2.2cm,0.7cm) -- +(-2.2cm,0.8cm);
\draw [blue,<->] (filter4.north west) +(-2.2cm,0.9cm) -- +(-0.8cm,0.9cm) node[anchor=south, midway] {$\frac{T}{2}$};
\end{tikzpicture}
\caption{}
\end{subfigure}
\begin{subfigure}{\textwidth}
\centering
%
%
%
\begin{tikzpicture}

\node[block] (filter1) {$q'(t) e^{-j\frac{\pi}{T}t}$};
\node[block] (filter2) [below=of filter1] {$q'(t) e^{-j\frac{\pi}{T}t}$};
\node[block] (filter3) [below=of filter2] {$q'(t) e^{-j\frac{\pi}{T}t}$};
\node[block] (filter4) [below=of filter3, yshift=-1.2cm] {$q'(t) e^{-j\frac{\pi}{T}t}$};

\node[circle, fill=black, below=of filter3, yshift=-0.0cm, inner sep=0pt, minimum width=0.1cm] {};
\node[circle, fill=black, below=of filter3, yshift=-0.2cm, inner sep=0pt, minimum width=0.1cm] {};
\node[circle, fill=black, below=of filter3, yshift=-0.4cm, inner sep=0pt, minimum width=0.1cm] {};

\foreach \x in {1,2,3,4}
	\node(Re\x) [block, right of=filter\x, xshift=1cm] {Re};
	
\foreach \x in {1,2,3,4}
	\draw [->] (filter\x) -- (Re\x);

\node (phasemul2) [summul,left of=filter2, xshift=-1cm] {\large $\times$};
\node (phasemul4) [summul,left of=filter4, xshift=-1cm] {\large $\times$};

\foreach \x in {2,3,4}
	\node (mul\x) [summul,left of=filter\x, xshift=-2.4cm] {\large $\times$};

\draw [<-] (filter2) -- (phasemul2);
\draw [<-] (phasemul2) -- (mul2);
\draw [<-] (filter3) -- (mul3);
\draw [<-] (filter4) -- (phasemul4);
\draw [<-] (phasemul4) -- (mul4);
	
\draw [<-] (mul2) -- +(0cm,0.8cm) node [anchor=south] {$e^{-j \frac{2\pi}{T}t}$};
\draw [<-] (mul3) -- +(0cm,0.8cm) node [anchor=south] {$e^{-j 2\frac{2\pi}{T}t}$};
\draw [<-] (mul4) -- +(0cm,0.8cm) node [anchor=south] {$e^{-j (N-1)\frac{2\pi}{T}t}$};

\draw [<-] (phasemul2) -- +(0cm,0.8cm) node [anchor=south] {$e^{-j \frac{\pi}{2}}$};
\draw [<-] (phasemul4) -- +(0cm,0.8cm) node [anchor=south] {$e^{-j \frac{\pi}{2}}$};

\foreach \x/\n in {1/0,2/1,3/2,4/(N-1)}
{
	\draw [->] (Re\x.east) -- +(0.5cm,0) node [anchor=north west] {$\frac{T}{2}$};
	\draw (Re\x.east) +(0.5cm,0) -- +(0.8cm,0.4cm);
	\draw [->] (Re\x.east) +(0.5cm,0.4) to [bend left=30] +(0.8cm,0cm);
	\draw (Re\x.east) +(0.5cm,0) +(0.8cm,0cm) -- +(1.5cm,0cm) node [anchor=south] {$\hat{A}^I_{k,\n}$} -- +(1.5cm,0.1cm) ;
	\draw (Re\x.east) +(1.5cm,0) -- +(2.9cm,0cm) [->] node [anchor=south] {$\hat{A}^R_{k,\n}$};
}

\node (input) [left of=filter2, xshift=-4.5cm,yshift=-1cm ,label=above:$s(t)$] {} ;
\draw (input) -- +(1cm,0);
\draw (input) +(1cm,0) |- (filter1.west) [->];
\draw (input) +(1cm,0) |- (mul2.west) [->];
\draw (input) +(1cm,0) |- (mul3.west) [->];
\draw (input) +(1cm,0) |- (mul4.west) [->];

\end{tikzpicture}
\caption{}
\end{subfigure}
\caption{The CMT (a) modulator and (b) demodulator, for QAM data symbols and zero-phase $q'(t)$.}
\label{fig:CMT-QAM-mod-demod}
\end{figure}
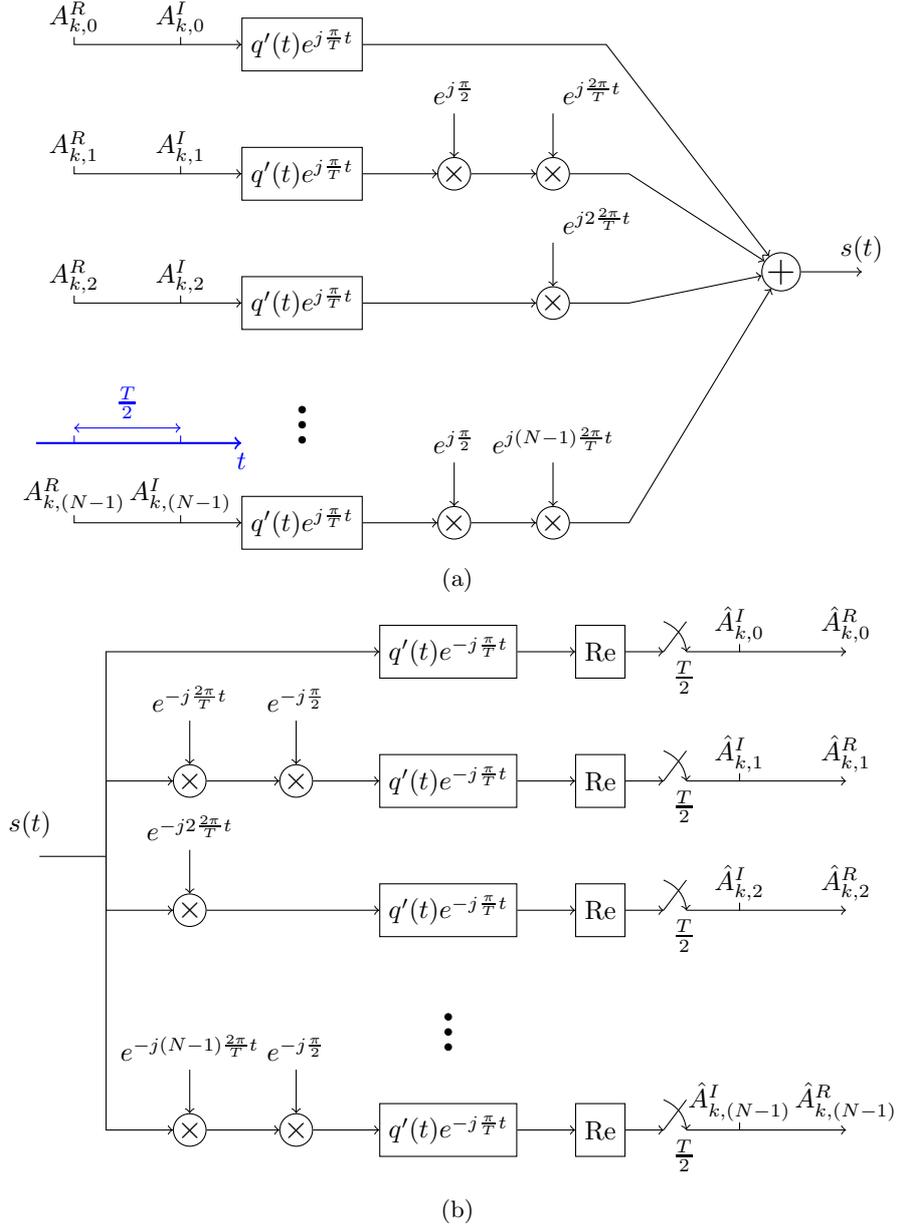

This provides an interesting insight into this development. That is, using double side-band real pulses, as the key idea behind the CMT scheme, is equivalent to applying $(-j)^k$ to the data symbols. Although it takes the values from $\{1,-j,-1,j\}$ periodically, only the $\frac{\pi}{2}$ phase difference among the adjacent subchannels is important. Therefore, a factor alternating between $1$ and $j$ is enough. \saeed{not solid enough? elaborate?}

\begin{figure}[!t]%
\centering
%
%
\begin{tikzpicture}
\draw [axis-line] (-4,0) -- (6,0) node[below] {$t$};
\draw [axis-line] (-4,-3) -- (6,-3) node[below] {$t$};

\foreach \x in {2,...,4}
{
	\draw [axis-tick] (\x,0) -- +(0,-0.1) node[below] {$\x T$};
	\draw [axis-tick] (\x,-3) -- +(0,-0.1) node[below] {$\x T$};
}
\draw [axis-tick] (1,0) -- +(0,-0.1) node[below] {$T$};
\draw [axis-tick] (1,-3) -- +(0,-0.1) node[below] {$T$};
\draw [axis-tick] (-1,0) -- +(0,-0.1) node[below] {$-T$};
\draw [axis-tick] (-1,-3) -- +(0,-0.1) node[below] {$-T$};
\draw [axis-tick] (0,-3) -- +(0,-0.1) node[below] {$0$};
\draw [axis-tick] (0,0) -- +(0,-0.1) node[below] {$0$};

\foreach \x in {2,...,3}
{
	\draw [axis-tick] (-\x,0) -- +(0,-0.1) node[below] {$-\x T$};
	\draw [axis-tick] (-\x,-3) -- +(0,-0.1) node[below] {$-\x T$};
}


\draw [blue] (-3,0) .. controls +(0.2,0.2) and +(-0.2,0.2) .. +(1,0)
			.. controls +(0.2,-0.4) and +(-0.1,-0.4) .. +(2,0)
			.. controls +(0.1,0.5) and +(-0.5,0) ..+(3,2)
			.. controls +(0.5,0) and +(-0.1,0.5) .. +(4,0)
			.. controls +(0.1,-0.4) and +(-0.2,-0.4) .. +(5,0)
			.. controls +(0.1,0.2) and +(-0.2,0.2) .. +(6,0);

\draw [black] (-2,0) .. controls +(0.2,0.2) and +(-0.2,0.2) .. +(1,0)
			.. controls +(0.2,-0.4) and +(-0.1,-0.4) .. +(2,0)
			.. controls +(0.1,0.5) and +(-0.5,0) ..+(3,2)
			.. controls +(0.5,0) and +(-0.1,0.5) .. +(4,0)
			.. controls +(0.1,-0.4) and +(-0.2,-0.4) .. +(5,0)
			.. controls +(0.1,0.2) and +(-0.2,0.2) .. +(6,0);

\draw [red] (-1,0) .. controls +(0.2,0.2) and +(-0.2,0.2) .. +(1,0)
			.. controls +(0.2,-0.4) and +(-0.1,-0.4) .. +(2,0)
			.. controls +(0.1,0.5) and +(-0.5,0) ..+(3,2)
			.. controls +(0.5,0) and +(-0.1,0.5) .. +(4,0)
			.. controls +(0.1,-0.4) and +(-0.2,-0.4) .. +(5,0)
			.. controls +(0.1,0.2) and +(-0.2,0.2) .. +(6,0);

\draw [blue] (-3,-5) -- +(+6,+4);
\draw [black] (-2,-5) -- +(+6,+4);
\draw [red] (-1,-5) -- +(+6,+4);

\draw [black, <->] (2,-1.7) -- +(0,-0.6) node [right, midway, xshift=-0.1cm] {\small$\frac{\pi}{2}$};
\draw [black, <->] (3,-1.7) -- +(0,-0.6) node [right, midway, xshift=-0.1cm] {\small$\frac{\pi}{2}$};
\draw [black] (3.8,-1.83) -- +(+1,0) node [above, xshift=0.5cm]{\small slope=$\frac{\pi}{2T}$};

\draw (-2,1.5) node {$q(t-kT)$};
\draw (-2,-2) node {$e^{j\! \frac{\pi}{2T}(t-kT)}$};

\end{tikzpicture}
%
%
%
\caption{One implication of the phase rotation applied to $q(t)$ as in \eqn{eq:CMT-pulses-time-single-side}.}%
\label{fig:CMT-phase}%
\end{figure}

\begin{figure}[!t]%
\centering
\begin{subfigure}{\textwidth}
\centering
%
%
%
\begin{tikzpicture}

\node[block] (filter1) {$q'(t)$};
\node[block] (filter2) [below=of filter1] {$q'(t)$};
\node[block] (filter3) [below=of filter2] {$q'(t)$};
\node[block] (filter4) [below=of filter3, yshift=-1.2cm] {$q'(t)$};

\node[circle, fill=black, below=of filter3, yshift=-0.0cm, inner sep=0pt, minimum width=0.1cm] {};
\node[circle, fill=black, below=of filter3, yshift=-0.2cm, inner sep=0pt, minimum width=0.1cm] {};
\node[circle, fill=black, below=of filter3, yshift=-0.4cm, inner sep=0pt, minimum width=0.1cm] {};

\node (phasemul2) [summul,right of=filter2, xshift=1cm] {\large $\times$};
\node (phasemul4) [summul,right of=filter4, xshift=1cm] {\large $\times$};

\foreach \x in {2,3,4}
	\node (mul\x) [summul,right of=filter\x, xshift=2.3cm] {\large $\times$};
	
\draw [->] (filter2) -- (phasemul2);
\draw [->] (filter3) -- (mul3);
\draw [->] (filter4) -- (phasemul4);
\draw [->] (phasemul2) -- (mul2);
\draw [->] (phasemul4) -- (mul4);
	
\foreach \x/\n in {1/0,2/1,3/2, 4/(N-1)}
{
	\draw [<-] (filter\x.west) -- +(-0.8cm,0) node [anchor=south] {$j A^I_{k,\n}$}
	-- +(-0.8cm,0.1cm);
	\draw (filter\x.west) +(-0.8cm,0) -- +(-2.3cm,0) node [anchor=south] {$A^R_{k,\n}$}
	-- +(-2.3cm,0.1cm);
}

\draw [<-] (mul2) -- +(0cm,0.8cm) node [anchor=south west] {$e^{j \frac{2\pi}{T}t}$};
\draw [<-] (mul3) -- +(0cm,0.8cm) node [anchor=south west] {$e^{j 2\frac{2\pi}{T}t}$};
\draw [<-] (mul4) -- +(0cm,0.8cm) node [anchor=south] {$e^{j (N-1)\frac{2\pi}{T}t}$};

\draw [<-] (phasemul2) -- +(0cm,0.8cm) node [anchor=south] {$e^{j \frac{\pi}{2}}$};
\draw [<-] (phasemul4) -- +(0cm,0.8cm) node [anchor=south] {$e^{j \frac{\pi}{2}}$};

\node (adder) [summul,right of=mul2, xshift=2cm, yshift=-1.3cm] {\Large +};

\draw [->] (filter1) -- +(4cm,0) -- (adder);
\foreach \x in {2,3,4}
	\draw [->] (mul\x) -- +(1cm,0) -- (adder);

\node (extramul) [summul, right=of adder, xshift=-0.5cm] {\large $\times$};
\draw [->] (extramul.east) -- +(0.8cm,0) node [anchor=south] {$s(t)$};
\draw [->] (adder) -- (extramul);
\draw [<-] (extramul) -- +(0cm,0.8cm) node [anchor=south] {$e^{j\frac{\pi}{T}}$};

\draw [<-,thick,blue] (filter4.north west) +(0,0.7cm) node[anchor=north] {$t$} -- +(-2.7cm,0.7cm);
\draw [blue] (filter4.north west) +(-0.8cm,0.7cm) -- +(-0.8cm,0.8cm);
\draw [blue] (filter4.north west) +(-2.3cm,0.7cm) -- +(-2.3cm,0.8cm);
\draw [blue,<->] (filter4.north west) +(-2.3cm,0.9cm) -- +(-0.8cm,0.9cm) node[anchor=south, midway] {$\frac{T}{2}$};
\end{tikzpicture}
\caption{}
\end{subfigure}
\begin{subfigure}{\textwidth}
\centering
%
%
%
\begin{tikzpicture}[scale=1]

\node[block] (filter1) {$q'(t)$};
\node[block] (filter2) [below=of filter1, yshift=-0.6cm] {$q'(t) $};
\node[block] (filter3) [below=of filter2, yshift=-0.6cm] {$q'(t) $};
\node[block] (filter4) [below=of filter3, yshift=-1.2cm] {$q'(t) $};

\node[circle, fill=black, below=of filter3, yshift=-0.0cm, inner sep=0pt, minimum width=0.1cm] {};
\node[circle, fill=black, below=of filter3, yshift=-0.2cm, inner sep=0pt, minimum width=0.1cm] {};
\node[circle, fill=black, below=of filter3, yshift=-0.4cm, inner sep=0pt, minimum width=0.1cm] {};

\foreach \x in {1,2,3,4}
{
	\node(Re\x) [block, right of=filter\x, xshift=3cm,yshift=0.6cm] {Re};
	\node(Im\x) [block, right of=filter\x, xshift=3cm,yshift=-0.6cm] {Im};
}

\foreach \x/\n in {1/0,2/1,3/2,4/(N-1)}
{
	\draw (filter\x.east) -- +(1.5cm,0);
	\fill (filter\x.east) +(1.5cm,0) circle (0.05cm);

	\draw (Re\x.west) -- +(-0.5cm,0);
	\fill (Re\x.west) +(-0.5cm,0) circle (0.05cm);	

	\draw (Im\x.west) -- +(-0.5cm,0);
	\fill (Im\x.west) +(-0.5cm,0) circle (0.05cm);

	\draw [->] (filter\x.east) +(1.5cm,0) -- +(2.1cm,0.55cm) ;

	\draw [->] (Re\x.east) -- +(1cm,0) node [anchor=south] {$\hat{A}_{k,\n}^R$};
	\draw [->] (Im\x.east) -- +(1cm,0) node [anchor=south] {$\hat{A}_{k,\n}^I$};
}

\fill (filter1.east) +(0.5cm,0) node [anchor=south, yshift=0.3cm, xshift=0.7cm] {\footnotesize $t=2m\frac{T}{2}$};
\fill (filter1.east) +(0.5cm,0) node [anchor=north, yshift=-0.3cm,xshift=0.5cm] {\footnotesize $t=(2m\!+\!1)\frac{T}{2}$};

\node (phasemul2) [summul,left of=filter2, xshift=-1cm] {\large $\times$};
\node (phasemul4) [summul,left of=filter4, xshift=-1cm] {\large $\times$};

\foreach \x in {2,3,4}
	\node (mul\x) [summul,left of=filter\x, xshift=-2.4cm] {\large $\times$};

\draw [<-] (filter2) -- (phasemul2);
\draw [<-] (phasemul2) -- (mul2);
\draw [<-] (filter3) -- (mul3);
\draw [<-] (filter4) -- (phasemul4);
\draw [<-] (phasemul4) -- (mul4);
	
\draw [<-] (mul2) -- +(0cm,0.8cm) node [anchor=south] {$e^{-j \frac{2\pi}{T}t}$};
\draw [<-] (mul3) -- +(0cm,0.8cm) node [anchor=south] {$e^{-j 2\frac{2\pi}{T}t}$};
\draw [<-] (mul4) -- +(0cm,0.8cm) node [anchor=south] {$e^{-j (N-1)\frac{2\pi}{T}t}$};

\draw [<-] (phasemul2) -- +(0cm,0.8cm) node [anchor=south] {$e^{-j \frac{\pi}{2}}$};
\draw [<-] (phasemul4) -- +(0cm,0.8cm) node [anchor=south] {$e^{-j \frac{\pi}{2}}$};


\node (extramul) [summul,left of=filter2, xshift=-4.5cm,yshift=-2cm] {\large $\times$};
\draw [<-] (extramul) -- +(0cm,0.8cm) node [anchor=south] {$e^{-j \frac{\pi}{T}t}$};

\draw [<-] (extramul) -- +(-1cm,0) node [anchor=south] {$\hat{s}(t)$};
\draw (extramul) -- +(1cm,0);
\draw (extramul)  +(1cm,0) |- (filter1.west) [->];
\draw (extramul) +(1cm,0) |- (mul2.west) [->];
\draw (extramul) +(1cm,0) |- (mul3.west) [->];
\draw (extramul) +(1cm,0) |- (mul4.west) [->];

\end{tikzpicture}
\caption{}
\end{subfigure}
\caption{The modified CMT modulator and demodulator.}
\label{fig:CMT-QAM1-mod-demod}
\end{figure}
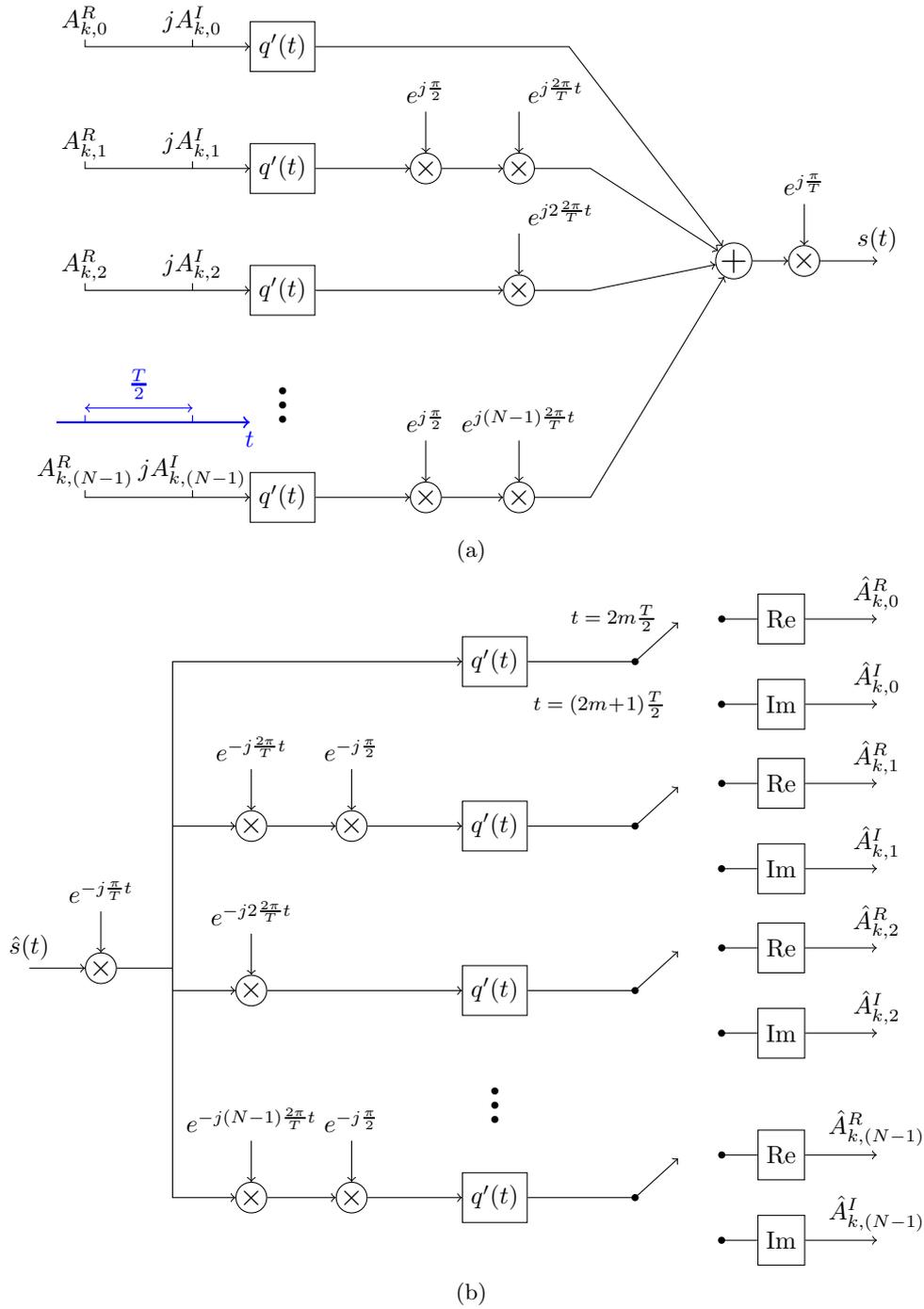

Therefore, the structure shown in \fig{fig:CMT-QAM-mod-demod} can be modified to obtain the one shown in \fig{fig:CMT-QAM1-mod-demod}. In this transformation, the derivation indicated in \eqn{eq:CMT-phase-interp} is used, such that the imaginary part of the QAM symbols are now purely imaginary. In the previous structure, the effect of the phase term in the pulses were compensated by its matched filter counterpart in the receiver. In the modified scheme, the compensation is done by using the $\im{.}$ operator instead of $\re{.}$ for the sampling instances when a purely imaginary symbol is expected. This is equivalent to a multiplication by the suitable scalar and then using the $\re{.}$ operator. Therefore, the working principle is unchanged.

In addition to the purely imaginary symbols, \eqn{eq:CMT-phase-interp} requires multiplication of the phase term $e^{j \frac{\pi}{2T}t}$ to the time-domain baseband signal. This is equivalent to a frequency shift. However, it is immediately canceled in the receiver before the $\re{.}$ or $\im{.}$ operators. Therefore, they can be omitted without violating the Nyquist criteria.

\saeedlong{So I'm saying that the small frequency shift, or the 1/2 factor in cosine term, which was the only reason these pulses (those double-sided ones of Chang) satisfy the Nyquist criteria, is now translated into pi/2 phase difference between adjacent symbols in time and a frequency shift on the whole signal which can be removed. So can't we reconstruct the whole thing from scratch using this idea? Though, it's probably not easy to represent problem-solving steps without surprising the reader!}

\begin{figure}[!t]
\centering
%
%
%
\begin{tikzpicture}

\node[block] (filterA1) [text width=1.8cm] {$q'(t)$};
\node[block] (filterB1) [below=of filterA1, yshift=0.6cm,text width=1.8cm] {$q'(t+\frac{T}{2})$};
\node[block] (filterA2) [below=of filterB1,text width=1.8cm] {$q'(t)$};
\node[block] (filterB2) [below=of filterA2, yshift=0.6cm,text width=1.8cm] {$q'(t+\frac{T}{2})$};
\node[block] (filterA3) [below=of filterB2, yshift=-1.5cm,text width=1.8cm] {$q'(t)$};
\node[block] (filterB3) [below=of filterA3, yshift=0.6cm,text width=1.8cm] {$q'(t+\frac{T}{2})$};

\node[circle, fill=black, below=of filterB2, yshift=-0cm, inner sep=0pt, minimum width=0.1cm] {};
\node[circle, fill=black, below=of filterB2, yshift=-0.2cm, inner sep=0pt, minimum width=0.1cm] {};
\node[circle, fill=black, below=of filterB2, yshift=-0.4cm, inner sep=0pt, minimum width=0.1cm] {};

\foreach \x in {1,2,3}
{
	\node(Re\x) [block, left of=filterA\x, xshift=-1cm] {Re};
	\node(Im\x) [block, left of=filterB\x, xshift=-1cm] {Im};
	
	\draw [<-] (filterA\x) -- (Re\x);
	\draw [<-] (filterB\x) -- (Im\x);
}	

\foreach \x in {1,2,3}
{
	\node (ReImConn\x) [left of=Re\x, yshift=-0.7cm, inner sep=0] {};
	\draw  [<-] (Re\x) -| (ReImConn\x.center);
	\draw  [<-] (Im\x) -| (ReImConn\x.center);
}

\foreach \x in {2,3}
{
	\node (phasemul\x) [summul,left of=ReImConn\x, xshift=0cm] {\large $\times$};
	\draw [->] (phasemul\x) -- (ReImConn\x);
}

\foreach \x in {2,3}
{
	\node (mul\x) [summul,left of=phasemul\x, xshift=-0.3cm] {\large $\times$};
	\draw [->] (mul\x) -- (phasemul\x);
}
	
\foreach \x/\n in {1/0,2/1,3/(N-1)}
{
	\draw [->] (filterA\x.east) -- +(0.5cm,0) node [anchor=north west] {\small $\frac{T}{2}$};
	\draw (filterA\x.east) +(0.5cm,0) -- +(0.8cm,0.4cm);
	\draw [->] (filterA\x.east) +(0.5cm,0.4) to [bend left=30] +(0.8cm,0cm);
	\draw (filterA\x.east) +(0.5cm,0) +(0.8cm,0cm) -- +(1.4cm,0cm) [->] node [anchor=south] {$\hat{A}^R_{\n}$};
	
	\draw [->] (filterB\x.east) -- +(0.5cm,0) node [anchor=north west] {\small $\frac{T}{2}$};
	\draw (filterB\x.east) +(0.5cm,0) -- +(0.8cm,0.4cm);
	\draw [->] (filterB\x.east) +(0.5cm,0.4) to [bend left=30] +(0.8cm,0cm);
	\draw (filterB\x.east) +(0.5cm,0) +(0.8cm,0cm) -- +(1.4cm,0cm) [->] node [anchor=south] {$\hat{A}^I_{\n}$};
}

\draw [<-] (mul2) -- +(0cm,0.8cm) node [anchor=south] {$e^{-j \frac{2\pi}{T}t}$};
\draw [<-] (mul3) -- +(0cm,0.8cm) node [anchor=south] {$e^{-j (N-1)\frac{2\pi}{T}t}$};

\draw [<-] (phasemul2) -- +(0cm,0.8cm) node [anchor=south] {$e^{-j \frac{\pi}{2}}$};
\draw [<-] (phasemul3) -- +(0cm,0.8cm) node [anchor=south] {$e^{-j \frac{\pi}{2}}$};

\node (input) [left of=mul2, xshift=-1cm,yshift=-1cm ,label=above:$\hat{s}(t)$] {} ;
\draw (input) -- +(1cm,0);
\draw (input) +(1cm,0) |- (ReImConn1.west) [->];
\draw (input) +(1cm,0) |- (mul2.west) [->];
\draw (input) +(1cm,0) |- (mul3.west) [->];

\end{tikzpicture}
\caption{The SMT demodulator.}
\label{fig:SMT-demod}
\end{figure}
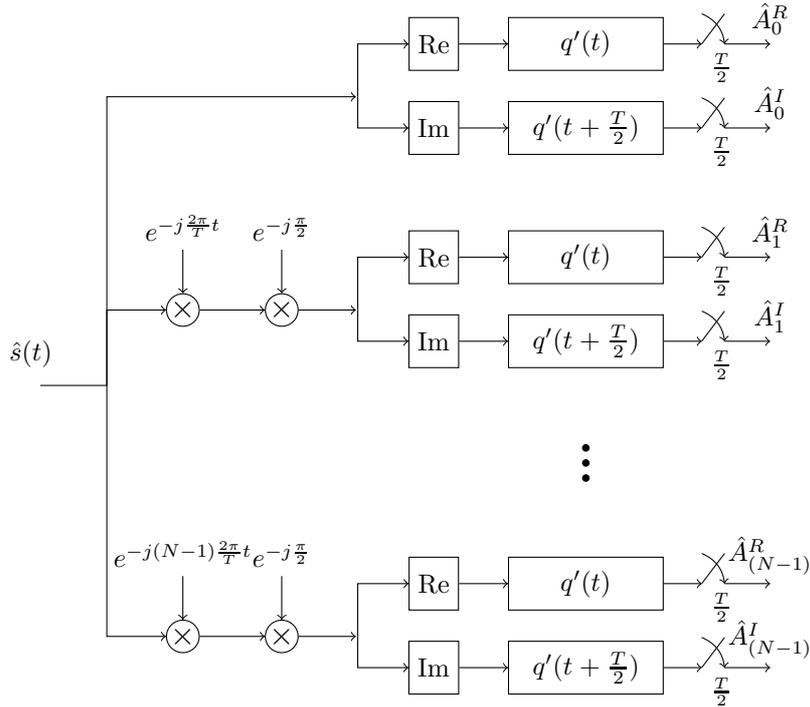

The structure shown in \fig{fig:CMT-QAM1-mod-demod} emphasizes the idea of splitting the QAM symbols into their real and imaginary parts. That is, the input symbols of the modulator are shown at consecutive time instants. Similarly, the output symbols of the demodulator are produced based on timing of the switch. However, a more familiar appearance of this scheme, commonly known as SMT, is shown in \fig{fig:SMT-mod} and \fig{fig:SMT-demod} \cite{farhangCMT}. In this structure, the subchannel filters that process the split symbols have a time offset of $\frac{T}{2}$, implying that symbols are processed consecutively in time. The receiver filters accordingly have a time offset. Note that in the receiver, the $\re{.}$ and $\im{.}$ operators are commuted with the filters. This is valid as long as the filters are real, which is the case in this development. 

\begin{figure}[!t]
\centering
%
%
%
\begin{tikzpicture}

\node[block] (filterA1) [text width=1.8cm] {$q'(t)$};
\node[block] (filterB1) [below=of filterA1, yshift=0.6cm,text width=1.8cm] {$q'(t-\frac{T}{2})$};
\node[block] (filterA2) [below=of filterB1,text width=1.8cm] {$q'(t)$};
\node[block] (filterB2) [below=of filterA2, yshift=0.6cm,text width=1.8cm] {$q'(t-\frac{T}{2})$};
\node[block] (filterA3) [below=of filterB2, yshift=-1.5cm,text width=1.8cm] {$q'(t)$};
\node[block] (filterB3) [below=of filterA3, yshift=0.6cm,text width=1.8cm] {$q'(t-\frac{T}{2})$};

\node[circle, fill=black, below=of filterB2, yshift=-0cm, inner sep=0pt, minimum width=0.1cm] {};
\node[circle, fill=black, below=of filterB2, yshift=-0.2cm, inner sep=0pt, minimum width=0.1cm] {};
\node[circle, fill=black, below=of filterB2, yshift=-0.4cm, inner sep=0pt, minimum width=0.1cm] {};

\foreach \x in {1,2,3}
{
	\node (filter-adder\x) [summul, right=of filterA\x, yshift=-0.7cm] {\Large +};
	\draw [->] (filterA\x) -| (filter-adder\x);
	\draw [->] (filterB\x.east) -| (filter-adder\x) ;
}

\foreach \x in {2,3}
{
	\node (phasemul\x) [summul,right of=filter-adder\x, xshift=0cm] {\large $\times$};
	\draw [->] (filter-adder\x) -- (phasemul\x);
}

\foreach \x in {2,3}
{
	\node (mul\x) [summul,right of=phasemul\x, xshift=0cm] {\large $\times$};
	\draw [->] (phasemul\x) -- (mul\x);
}

\foreach \x/\n in {1/0,2/1,3/(N-1)}
{
	\draw [<-](filterA\x.west) -- +(-0.8cm,0) node [anchor=south] {$A^R_{k,\n}$};
	\draw [<-](filterB\x.west) -- +(-0.8cm,0) node [anchor=south] {$j A^I_{k,\n}$};
}

\draw [<-] (mul2) -- +(0cm,0.8cm) node [anchor=south west] {$e^{j \frac{2\pi}{T}t}$};
\draw [<-] (mul3) -- +(0cm,0.8cm) node [anchor=south west] {$e^{j (N-1)\frac{2\pi}{T}t}$};

\draw [<-] (phasemul2) -- +(0cm,0.8cm) node [anchor=south] {$e^{j \frac{\pi}{2}}$};
\draw [<-] (phasemul3) -- +(0cm,0.8cm) node [anchor=south] {$e^{j \frac{\pi}{2}}$};

\node (adder) [summul,right of=mul2, xshift=2cm, yshift=-1.3cm] {\Large +};

\draw [->] (filter-adder1) -- +(3.3cm,0) -- (adder);
\foreach \x in {2,3}
	\draw [->] (mul\x) -- +(1.3cm,0) -- (adder);

\draw [->] (adder.east) -- +(0.8cm,0) node [anchor=south] {$s(t)$};

\end{tikzpicture}
\caption{The SMT modulator.}
\label{fig:SMT-mod}
\end{figure}
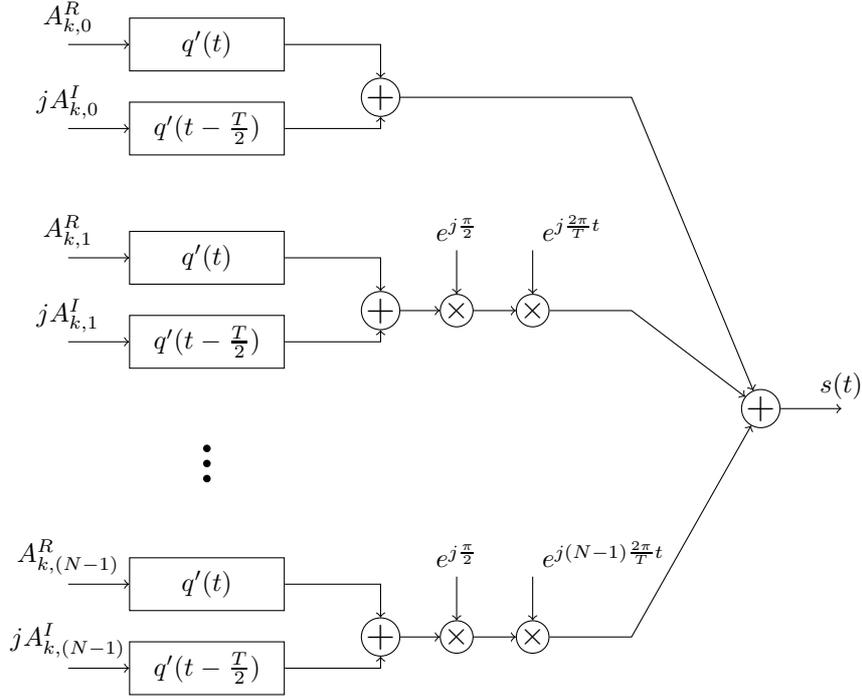

The $e^{j\frac{\pi}{2}}$ multipliers in the SMT modulator of \fig{fig:SMT-mod} can be translated to a scalar $j$ multiplied by the symbols. Thus, the QAM symbols are pre-processed before entering the modulator as shown in \fig{fig:OQAM-preprocessing-SMT}. It is easy to see that the modulation works correctly if there is a phase difference of $\frac{\pi}{2}$ between the adjacent symbols on one subcarrier and between the simultaneous symbols of the adjacent subcarriers. For instance, the implementation of the PHYDYAS project \cite{phydyas-D5.1}, which is used in this work as well, is based on the phase arrangement of \fig{fig:OQAM-preprocessing-PHYDYAS}. In addition, the order in which the real and imaginary parts of the QAM symbols are arranged may vary. Such variations do not change the mechanism of the modulation itself, but  modifies the pre- and post-processing of the QAM symbols into the so-called OQAM symbols and vice versa.

\begin{figure}[!t]%
\centering
\begin{subfigure}{0.4\textwidth}
\centering
%
%
%
\begin{tikzpicture}

\node (filter0) {};
\node (filter1) [below=of filter0, yshift=0.3cm] {};
\node (filter2) [below=of filter1, yshift=0.3cm] {};
\node (filter3) [below=of filter2, yshift=0.3cm] {};

\draw (filter0.west) -- +(-1cm,0) node [anchor=south] {$j A^I_{2,0}$} 
	-- +(-1cm,0.1cm) +(-1cm,0cm)
	-- +(-2.5cm,0) node [anchor=south] {$A^R_{2,0}$} 
	-- +(-2.5cm,0.1cm) +(-2.5cm,0cm)
	-- +(-4cm,0) node [anchor=south] {$jA^I_{1,0}$} 
	-- +(-4cm,0.1cm) +(-4cm,0cm)
	-- +(-5.5cm,0) node [anchor=south] {$A^R_{1,0}$} 
	-- +(-5.5cm,0.1cm) +(-5.5cm,0cm);

\draw (filter1.west) -- +(-1cm,0) node [anchor=south] {$- A^I_{2,1}$} 
	-- +(-1cm,0.1cm) +(-1cm,0cm)
	-- +(-2.5cm,0) node [anchor=south] {$jA^R_{2,1}$} 
	-- +(-2.5cm,0.1cm) +(-2.5cm,0cm)
	-- +(-4cm,0) node [anchor=south] {$-A^I_{1,1}$} 
	-- +(-4cm,0.1cm) +(-4cm,0cm)
	-- +(-5.5cm,0) node [anchor=south] {$jA^R_{1,1}$} 
	-- +(-5.5cm,0.1cm) +(-5.5cm,0cm);

\draw (filter2.west) -- +(-1cm,0) node [anchor=south] {$j A^I_{2,2}$} 
	-- +(-1cm,0.1cm) +(-1cm,0cm)
	-- +(-2.5cm,0) node [anchor=south] {$A^R_{2,2}$} 
	-- +(-2.5cm,0.1cm) +(-2.5cm,0cm)
	-- +(-4cm,0) node [anchor=south] {$jA^I_{1,2}$} 
	-- +(-4cm,0.1cm) +(-4cm,0cm)
	-- +(-5.5cm,0) node [anchor=south] {$A^R_{1,2}$} 
	-- +(-5.5cm,0.1cm) +(-5.5cm,0cm);

\draw (filter3.west) -- +(-1cm,0) node [anchor=south] {$- A^I_{2,3}$} 
	-- +(-1cm,0.1cm) +(-1cm,0cm)
	-- +(-2.5cm,0) node [anchor=south] {$jA^R_{2,3}$} 
	-- +(-2.5cm,0.1cm) +(-2.5cm,0cm)
	-- +(-4cm,0) node [anchor=south] {$-A^I_{1,3}$} 
	-- +(-4cm,0.1cm) +(-4cm,0cm)
	-- +(-5.5cm,0) node [anchor=south] {$jA^R_{1,3}$} 
	-- +(-5.5cm,0.1cm) +(-5.5cm,0cm);


\end{tikzpicture}
\caption{}
\label{fig:OQAM-preprocessing-SMT}
\end{subfigure}
\hspace{0.05\textwidth}
\begin{subfigure}{0.4\textwidth}
\centering
%
%
%
\begin{tikzpicture}

\node (filter0) {};
\node (filter1) [below=of filter0, yshift=0.3cm] {};
\node (filter2) [below=of filter1, yshift=0.3cm] {};
\node (filter3) [below=of filter2, yshift=0.3cm] {};

\draw (filter0.west) -- +(-1cm,0) node [anchor=south] {$-j A^I_{2,0}$} 
	-- +(-1cm,0.1cm) +(-1cm,0cm)
	-- +(-2.5cm,0) node [anchor=south] {$-A^R_{2,0}$} 
	-- +(-2.5cm,0.1cm) +(-2.5cm,0cm)
	-- +(-4cm,0) node [anchor=south] {$jA^I_{1,0}$} 
	-- +(-4cm,0.1cm) +(-4cm,0cm)
	-- +(-5.5cm,0) node [anchor=south] {$A^R_{1,0}$} 
	-- +(-5.5cm,0.1cm) +(-5.5cm,0cm);

\draw (filter1.west) -- +(-1cm,0) node [anchor=south] {$A^I_{2,1}$} 
	-- +(-1cm,0.1cm) +(-1cm,0cm)
	-- +(-2.5cm,0) node [anchor=south] {$-jA^R_{2,1}$} 
	-- +(-2.5cm,0.1cm) +(-2.5cm,0cm)
	-- +(-4cm,0) node [anchor=south] {$-A^I_{1,1}$} 
	-- +(-4cm,0.1cm) +(-4cm,0cm)
	-- +(-5.5cm,0) node [anchor=south] {$jA^R_{1,1}$} 
	-- +(-5.5cm,0.1cm) +(-5.5cm,0cm);

\draw (filter2.west) -- +(-1cm,0) node [anchor=south] {$j A^I_{2,2}$} 
	-- +(-1cm,0.1cm) +(-1cm,0cm)
	-- +(-2.5cm,0) node [anchor=south] {$A^R_{2,2}$} 
	-- +(-2.5cm,0.1cm) +(-2.5cm,0cm)
	-- +(-4cm,0) node [anchor=south] {$-jA^I_{1,2}$} 
	-- +(-4cm,0.1cm) +(-4cm,0cm)
	-- +(-5.5cm,0) node [anchor=south] {$-A^R_{1,2}$} 
	-- +(-5.5cm,0.1cm) +(-5.5cm,0cm);

\draw (filter3.west) -- +(-1cm,0) node [anchor=south] {$- A^I_{2,3}$} 
	-- +(-1cm,0.1cm) +(-1cm,0cm)
	-- +(-2.5cm,0) node [anchor=south] {$jA^R_{2,3}$} 
	-- +(-2.5cm,0.1cm) +(-2.5cm,0cm)
	-- +(-4cm,0) node [anchor=south] {$A^I_{1,3}$} 
	-- +(-4cm,0.1cm) +(-4cm,0cm)
	-- +(-5.5cm,0) node [anchor=south] {$-jA^R_{1,3}$} 
	-- +(-5.5cm,0.1cm) +(-5.5cm,0cm);


\end{tikzpicture}
\caption{}
\label{fig:OQAM-preprocessing-PHYDYAS}
\end{subfigure}
\caption{The preprocessing of the QAM symbols before entering the modulator, as in (a) SMT scheme, (b) implementation of the PHYDYAS project.}
\label{fig:OQAM-preprocessing}
\end{figure}

\section{Discrete-time model, OFDM/OQAM}
\label{sec:oqam}

This chapter is concluded by introducing a practical implementation of the developed SMT scheme. The details are based on the structure used in the PHYDYAS project~\cite{phydyas-D5.1}.
The SMT scheme is formalized and commonly used in the literature under the name  ``OFDM with offset OQAM'' or OFDM/OQAM. A discrete-time version  of this model is developed in \cite{995073}. It uses DFT-based modulation, as in OFDM, as well as the polyphase representation of the pulse shaping filter to achieve an efficient implementation of the OFDM/OQAM scheme.

The discrete-time model and its implementation are briefly reviewed here. The discussion here is incomplete and lacks continuity. The interested reader would find a detailed development in \cite{995073} and the required background in \cite{Vaidyanathan}.

\begin{figure}[!t]%
\centering
\begin{subfigure}{0.6\textwidth}
\centering
%
%
%
\begin{tikzpicture}


\node[block,minimum width=2cm] (filter1) {$A_0(z^2)$};
\node[block,minimum width=2cm] (filter2) [below=of filter1, yshift=-0.2cm] {$A_1(z^2)$};
\node[block,minimum width=2cm] (filter3) [below=of filter2, yshift=-2cm] {$A_{N-1}(z^2)$};

\node[block,minimum width=1.5cm, minimum height=6.5cm] (DFT) [left of=filter1, xshift=-1.5cm, yshift=-3cm] {IDFT};

\node[circle, fill=black, below=of filter2, yshift=-0.5cm, inner sep=0pt, minimum width=0.1cm] {};
\node[circle, fill=black, below=of filter2, yshift=-0.7cm, inner sep=0pt, minimum width=0.1cm] {};
\node[circle, fill=black, below=of filter2, yshift=-0.9cm, inner sep=0pt, minimum width=0.1cm] {};

\foreach \x in {1,2,3}
	\node (mul\x) [summul, left of=filter\x, xshift=-3cm] {$\times$};

\foreach \x/\n in {1/0,2/1,3/N-1}
	\draw [<-] (mul\x) -- +(0,0.5cm) node [anchor=south] {$\beta_{\n}$};

\foreach \x in {1,2,3}
{
	\draw [->] (mul\x.east) -- +(0.55cm,0) ;
	\draw [<-] (filter\x.west) -- +(-0.73cm,0) ;		
}

\foreach \x/\n in {1/0,2/1,3/N-1}
	\draw [<-] (mul\x.west) -- +(-0.8cm,0) node [anchor=south] {$A_{\n}$};

\foreach \x in {1,2,3}
{
	\node (upsampler\x) [block, right of=filter\x, xshift=1cm] {$\uparrow \frac{N}{2}$};
	\draw [->] (filter\x.east) -- (upsampler\x.west);
}

\node (delay1) [block, right of=upsampler1, xshift=0.5cm, yshift=-1.1cm] {$z^{-1}$};
\node (delay2) [block, right of=upsampler2, xshift=0.5cm, yshift=-1.2cm] {$z^{-1}$};
\node (delay3) [block, right of=upsampler3, xshift=0.5cm, yshift=0.8cm] {$z^{-1}$};

\node (adder1) [summul,right of=upsampler1, xshift=0.5cm, yshift=0cm] {+};
\node (adder2) [summul,right of=upsampler2, xshift=0.5cm, yshift=0cm] {+};

\draw [->] (upsampler1.east) -- (adder1);
\draw [->] (delay1) -- (adder1);

\draw [->] (upsampler2.east) -- (adder2);
\draw [->] (delay2) -- (adder2);
\draw [->] (adder2) -- (delay1);

\draw [->] (upsampler3.east) -| (delay3.south);

\draw [->] (delay3.north) -- +(0,0.3cm);
\draw [<-] (delay2.south) -- +(0,-0.3cm);

\node[circle, fill=black, below=of delay2, yshift=+0.6cm, inner sep=0pt, minimum width=0.05cm] {};
\node[circle, fill=black, below=of delay2, yshift=0.5cm, inner sep=0pt, minimum width=0.05cm] {};
\node[circle, fill=black, below=of delay2, yshift=0.4cm, inner sep=0pt, minimum width=0.05cm] {};

\draw [->] (adder1.east) -- +(0.8cm,0) node [anchor=south] {$s[k]$};

\end{tikzpicture}
\caption{}
\label{fig:SFB}
\end{subfigure}
\begin{subfigure}{0.6\textwidth}
\centering
%
%
%
\begin{tikzpicture}


\node[block,minimum width=2cm] (filter1) {$A_0(z^2)$};
\node[block,minimum width=2cm] (filter2) [below=of filter1, yshift=-0.2cm] {$A_1(z^2)$};
\node[block,minimum width=2cm] (filter3) [below=of filter2, yshift=-2cm] {$A_{N-1}(z^2)$};

\node[block,minimum width=1.5cm, minimum height=6.5cm] (DFT) [right of=filter1, xshift=+1.5cm, yshift=-3cm] {DFT};

\node[circle, fill=black, below=of filter2, yshift=-0.5cm, inner sep=0pt, minimum width=0.1cm] {};
\node[circle, fill=black, below=of filter2, yshift=-0.7cm, inner sep=0pt, minimum width=0.1cm] {};
\node[circle, fill=black, below=of filter2, yshift=-0.9cm, inner sep=0pt, minimum width=0.1cm] {};

\foreach \x in {1,2,3}
	\node (mul\x) [summul, right of=filter\x, xshift=+3cm] {$\times$};

\foreach \x/\n in {1/0,2/1,3/N-1}
	\draw [->] (mul\x) -- +(0,+0.5cm) node [anchor=south] {$\beta_{\n}$};

\foreach \x in {1,2,3}
{
	\draw [<-] (mul\x.west) -- +(-0.55cm,0) ;
	\draw [->] (filter\x.east) -- +(+0.73cm,0) ;		
}

\foreach \x/\n in {1/0,2/1,3/N-1}
	\draw [->] (mul\x.east) -- +(+0.8cm,0) node [anchor=south] {$\hat{A}_{\n}$};

\foreach \x in {1,2,3}
{
	\node (upsampler\x) [block, left of=filter\x, xshift=-1cm] {$\downarrow \frac{N}{2}$};
	\draw [<-] (filter\x.west) -- (upsampler\x.east);
}

\node (delay1) [block, left of=upsampler1, xshift=-0.5cm, yshift=-1.1cm] {$z^{-1}$};
\node (delay2) [block, left of=upsampler2, xshift=-0.5cm, yshift=-1.2cm] {$z^{-1}$};
\node (delay3) [block, left of=upsampler3, xshift=-0.5cm, yshift=0.8cm] {$z^{-1}$};

\draw [<->] (upsampler1) -| (delay1);

\draw [<-] (upsampler2) -| (delay2);
\draw [<-] (delay2) -- (delay1);

\draw [<-] (upsampler3.west) -| (delay3.south);

\draw [<-] (delay3.north) -- +(0,0.3cm);
\draw [->] (delay2.south) -- +(0,-0.3cm);

\node[circle, fill=black, below=of delay2, yshift=+0.6cm, inner sep=0pt, minimum width=0.05cm] {};
\node[circle, fill=black, below=of delay2, yshift=0.5cm, inner sep=0pt, minimum width=0.05cm] {};
\node[circle, fill=black, below=of delay2, yshift=0.4cm, inner sep=0pt, minimum width=0.05cm] {};

\draw [<-] (upsampler1.west) -- +(-1.8cm,0) node [anchor=south] {$s'[k]$};

\end{tikzpicture}
\caption{}
\label{fig:AFB}
\end{subfigure}
\caption{The structure of the efficient implementation of the OFDM/OQAM model. (a) The modulator or the Synthesis Filter Bank (SFB), (b) the demodulator or the Analysis Filter Bank (AFB).}
\label{fig:SFB-AFB}
\end{figure}
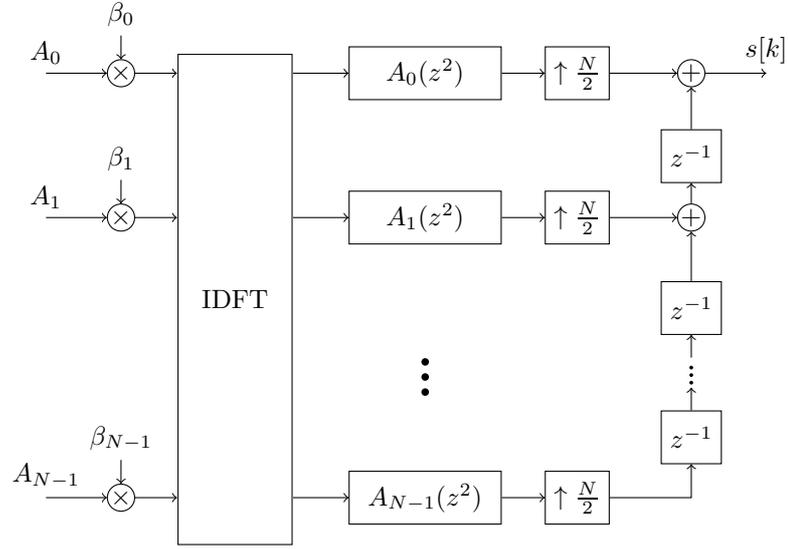
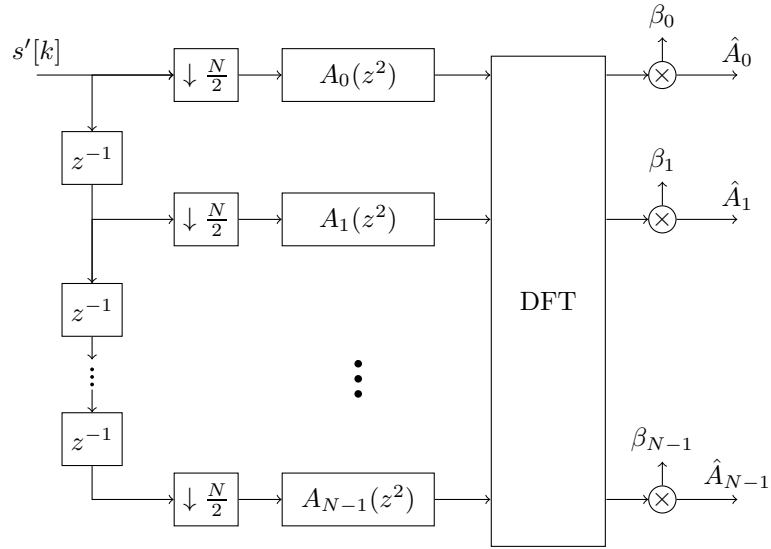

The discrete-time formulation of the OFDM/OQAM signal model requires causal prototype filter $h[k]$ which is the truncated and shifted version of the continuous-time filter $h(t)$. The final discrete-time signal model is \cite{995073}
\begin{equation}
s[k]=\sum_{m=-\infty}^{+\infty} \sum_{n=0}^{N-1} a_{m,n} h[k-m\frac{N}{2}] e^{j \frac{2\pi}{N} n (k-D/2)} e^{j\phi_{m,n}},
\label{eq:oqam-discrete}
\end{equation}
where the index $n$ refers to the subcarrier number and the index $m$ counts the intervals of $\frac{N}{2}$ samples which are transmitted for each set of $N$ OQAM symbols entering the modulator. Recall that each QAM symbol is split into two parts, referred to as OQAM symbols in this text. The length of the prototype filter $L_h$ is included in $D=L_h-1$. The real-valued real-valued symbols $a_{m,n}$ together with the phase term $e^{j\phi_{m,n}}$ form the OQAM symbols. The conversion of the QAM symbols to the OQAM ones is illustrated in \fig{fig:OQAM-preprocessing}. In the sequel, the efficient implementation of this model is discussed.

The modulator part, the Synthesis Filter Bank (SFB), is shown in \fig{fig:SFB}. The inputs to the SFB are the offset QAM symbols, as shown in \fig{fig:OQAM-preprocessing}. The IFFT block essentially performs the modulation to the subcarrier frequencies. From the hardware point of view, it performs the computations in a block processing manner. That is, a set of samples are fed into the $N$ branches of the IFFT block at once and a set of output samples are generated. After the polyphase filters, the upsampling by a factor of $N/2$ is performed. Through the specific combination of the delays and the adders, the resulting samples from the parallel branches go through a parallel-to-serial conversion. An important point is that for a set of samples at the input of IDFT, which are one part of the two-part QAM symbols, $\frac{N}{2}$ samples are put on the final output of the SFB. Viewing from one stage before, for each set of $N$ QAM symbols, $N$ samples are generated at the output. However, depending on the pulse shape, each OFDM/OQAM symbol is several times longer than $N$ samples. 

The demodulator part, the Analysis Filter Bank (AFB), is shown in \fig{fig:AFB}. The output of the parallel branches are the OQAM symbols which must go through OQAM-postprocessing which reverses the procedure shown in \fig{fig:OQAM-preprocessing}. \saeed{delay?}

\begin{figure}
	\centering
	\includegraphics[width=0.6\textwidth]{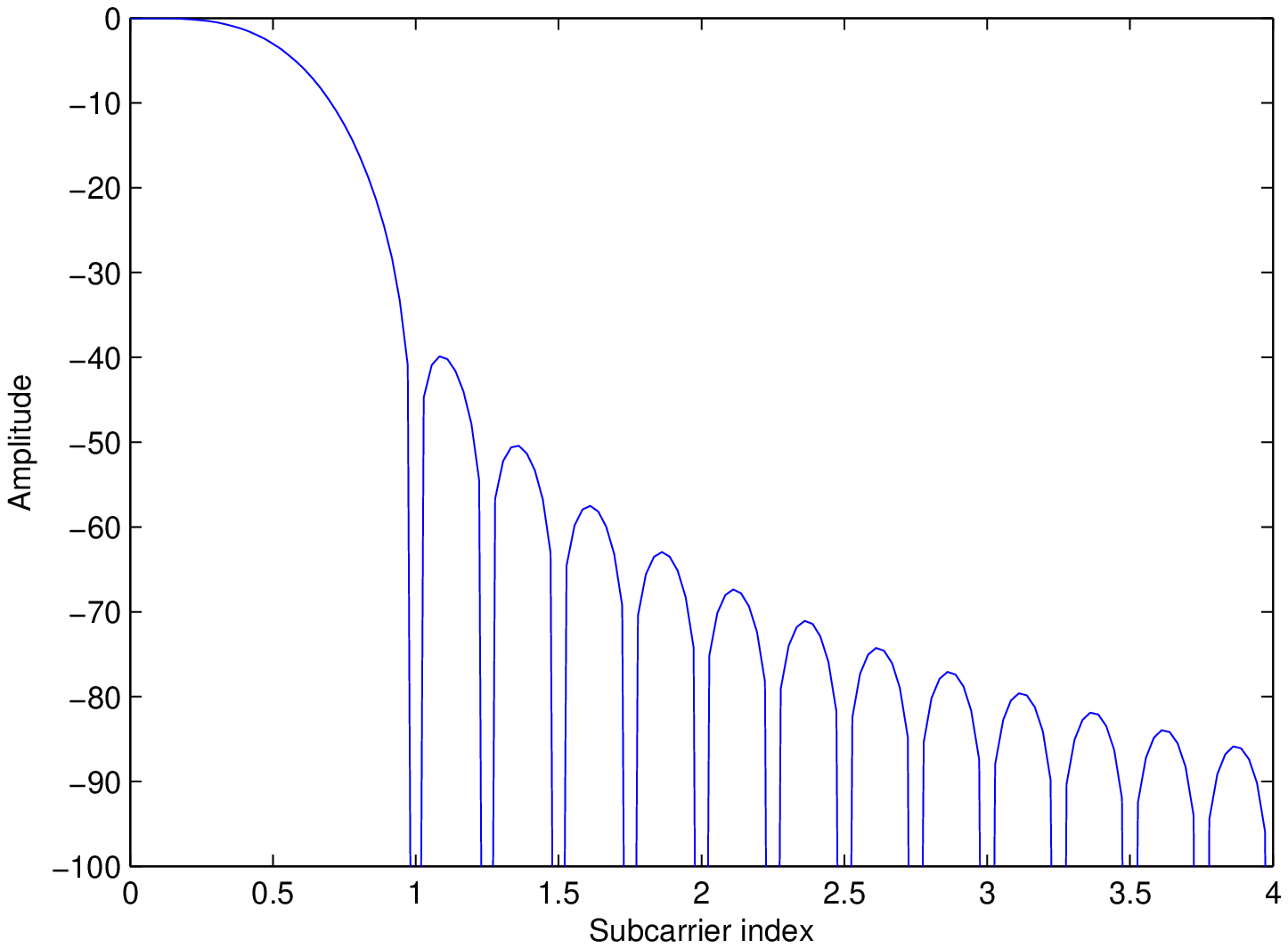}
	\caption{The pulse shape in the frequency domain. The frequency axis is normalized to the subcarrier spacing $\frac{2\pi}{N}$.}
	\label{fig:pulseShapeFreq}
\end{figure}

One important consideration is the pulse shape, namely the prototype filter, which satisfies the orthogonality condition, i.e., the discrete-time version of the Nyquist criterion. The pulse shape proposed in \cite{phydyas-D5.1} is presented here.  The impulse response of the causal FIR  prototype filter, denoted by $p[m]$, is defined as 
\begin{equation}
p[m]=\bar{P}[0]+2 \sum_{k=1}^{K-1} (-1)^k \bar{P}[k] \cos \left( \frac{2\pi k}{K\!M} (m+1)\right), \quad m=0,1,\ldots,K\!M-2,
\end{equation}
where $M$ is the number of the subcarriers, $K=4$ is the overlapping factor and the other constants are defined as
\begin{align*}
\bar{P}[0]&=1,\\
\bar{P}[1]&=0.97195983,\\
\bar{P}[2]&=\frac{1}{\sqrt{2}},\\
\bar{P}[3]&=\sqrt{1-\bar{P}[2]}.
\end{align*}
The prototype filter is designed for the fixed overlapping factor $K=4$. This indicates that the polyphase filters and the parallel-to-serial conversion in the SFB continue processing the samples of each symbol for the following 3 symbol intervals. \fig{fig:pulseShapeFreq} shows that the pulse shape in the frequency domain overlaps mainly over the adjacent channel, and has an attenuation of more than 60 dB afterwards.

It is evident that the discretization of the continuous-time model developed in the previous sections does not necessarily lead to perfect satisfaction of the orthogonality conditions. The described pulse shape is an example of such a design. The investigation of the details of this approximation is beyond the scope of this text. However, the mentioned literature provide an in-depth formalization of this issue and show that the interference due to the approximate orthogonality is negligible.

\clearpage
\bibliographystyle{plain}
\bibliography{Bibliography}

\end{document}